\documentclass{article}

\usepackage{amsmath}
\usepackage{graphicx}

\newcommand{\BOne}[5]{B_{#1 #2}\!\!\;(#3; #4, #5)}
\newcommand{\BOneNenner}[7]{B_{#1 #2;#4 #6}\!\!\;(#3; #5; #7)}
\newcommand{\ANenner}[2]{A_{0;#1}(#2)}
\newcommand{\vierpi}[1]{} 
\newcommand{\Diag}[4]{\mbox{$\Delta^{(#1)}_{#2} f_{#4}$}}
\newcommand{\Tr}{\textrm{Tr}}
\newcommand{\Lg}{\mbox{\rm Lg}}
\newcommand{\Lren}[1]{L_#1^{(0)}}

\begin{document}


\begin{flushright}
   MZ-TH--01-39  \\
   hep-ph/0112352\\
   December 2001 \\
\end{flushright}

\vspace*{1cm}

\begin{center}
\textbf{{\large \boldmath$K_{l3}$ form factors at order $p^{6}$ of chiral
perturbation theory }} \vspace{10mm} \\[0pt]
P.~Post\footnote{%
Now at Fraunhofer Gesellschaft, Institute for Algorithms and Scientific
Computing (SCAI), Schloss Birlinghoven, D-53754 Sankt Augustin, Germany,
E-mail: peter.post@scai.fraunhofer.de}
and K.~Schilcher\footnote{%
Supported in part by Bundesministerium f\"{u}r Bildung und Forschung, Bonn,
Germany, under Contract 05~HT9UMB~4, E-mail: karl.schilcher@uni-mainz.de}
\vspace{10mm} \\[0pt]
\emph{Institut f\"{u}r Physik, Johannes-Gutenberg-Universit\"{a}t,\\[0pt]
Staudinger Weg 7, D-55099 Mainz, Germany}
\end{center}

\bigskip

\textbf{Abstract: }
This paper describes the calculation of the semileptonic $K_{l3}$
decay form factors at order $p^6$ of chiral perturbation theory which is
the next-to-leading order correction to the well-known $p^4$ result
achieved by Gasser and Leutwyler. At order $p^6$ the chiral expansion
contains 1- and 2-loop diagrams which are discussed in detail. The
irreducible 2-loop graphs of the sunset topology are calculated
numerically. In addition, the chiral Lagrangian $\mathcal{L}^{(6)}$
produces direct couplings with the $W$-bosons. Due to these unknown
couplings, one can always add linear terms in $q^2$ to the predictions of
the form factor $f_-(q^2)$. For the form factor $f_+(q^2)$, this
ambiguity involves even quadratic terms. Making use of the fact that the
pion electromagnetic form factor involves the same $q^4$ counter term, the
$q^4$-ambiguity can be resolved. Apart from the possibility of adding an
arbitrary linear term in $q^2$ our calculation shows that chiral
perturbation theory converges very well in this application, as the
$\mathcal{O}(p^6)$-corrections are small. Comparing the predictions of
chiral perturbation theory with the recent CPLEAR data, it is seen that
the experimental form factor $f_+(q^2)$ is well described by a
linear fit, but that the slope $\lambda_+$ is  smaller by about 2 standard
deviations than the $\mathcal{O}(p^4)$ prediction.
The unavoidable $q^2$ counter term of the $\mathcal{O}(p^6)$
corrections allows to bring the predictions of chiral perturbation theory
into perfect agreement with experiment.

\vspace{12mm}

\newpage



\section{Introduction}

The hadronic matrix elements of weak decays constitute a decisive testing
ground for our understanding of low energy strong interactions. In this
respect, the semileptonic $K_{l3}$-decay is one of the cleanest and most
interesting processes. In particular, it has been stressed \cite%
{LeutwylerRoos} that this decay constitutes the best source for the
extraction of the CKM matrix element $\left| V_{us}\right|$. On the
experimental side there exist a number of old high statistics results (some
contradictory), and new precision experiments are in progress or already
published \cite{apostolakis}. On the theoretical side chiral perturbation
theory \cite{Weinberg79}\cite{GasserLeutw1} has established itself as a
powerful effective theory of low energy strong interactions. Based on the
symmetry of the underlying QCD, chiral perturbation theory produces a
systematic low-energy expansion of the observables in this regime.
Unfortunately, because of the non-renormalizability of the effective theory,
higher powers in the energy expansion require higher loop Feynman integrals
and as input an ever increasing number of renormalization constants. The
$p^{4}$-Lagrangian involves ten free parameters which were determined in the
fundamental papers of Gasser and Leutwyler \cite{GasserLeutw1}. 
In a more recent analysis \cite{BijLag} the number of independent counter
terms was reduced to 90. Unfortunately, there is no simple prescription how to
translate from one base into the other. The series of calculations on form
factors taken up by us begun before this paper was published. In order to
preserve the continuity with our previous papers on order $p^6$ chiral
perturbation theory, we stick to the convention of \cite{FearingScherer}. 
There is no harm in keeping more than the minimal number of operators, 
as it is, for all practical purposes, impossible to determine by the
necessary  
independent experiments such a large number of counterterms 
(including their finite parts), be it 143 or 90. 
Relations between individual physical processes are manifest 
in any case, as demonstrated in \cite{PostSchilcher1997} 
and in the discussion below, where a relation of a part of the $K_{l3}$ 
form factor to the pion electromagnetic form factor is established.
The $K_{l3}$-decay amplitude has been
calculated some time ago to $\mathcal{O}(p^{4})$ \cite{GasserLeutw2}
and more recently to $\mathcal{O}(p^{4},(m_{d}-m_{u})p^{2},e^{2}p^{2})$
\cite{neufeld}. 
To $\mathcal{O}(p^6)$, the contribution of the double chiral logs has been 
calculated in \cite{BijDoubleChLog}.

We present here the results of a full $p^{6}$ and two-loop analysis of the
semi-leptonic $K_{l3}$ form factors, relying heavily on recent progress 
in the calculation of massive two-loop integrals \cite{PostTausk}\cite{PostSchilcher}. 
Complementary calculations to this order involve the vector two-point function 
\cite{GoloKam} and the $K_{l4}$ form factors \cite{AmorosBij}.

As the relevant part of the $p^{6}$-Lagrangian contains so many unknown 
parameters, one may question the usefulness of such calculations. 
Here are a few arguments in favour:

\begin{itemize}
\item In a given class of experiments, such as the electromagnetic and weak
form factors of the light mesons, only a limited number of renormalization
constants enter and relations between amplitudes can be tested \cite%
{PostSchilcher1997}.

\item The unknown constants enter only polynomially, and precision
experiments could separate the unambiguous predictions.

\item Knowledge of the exact low-energy functional form of an amplitude may
be important for the experimental extraction of low-energy parameters such
as charge radii.

\item The results may be used in model calculations which predict the
polynomial terms. These calculations can then be compared with experiment.

\item The question of convergence of the chiral perturbation theory may be
addressed.
\end{itemize}


\section{The matrix element}

\bigskip $K$-mesons can decay into a pion and a lepton pair via the
following channels
\begin{eqnarray}
K^{+}(p_{1}) &\rightarrow& \pi^{0}(p_{2}) \: \ell^{+}(p_{\ell}) \: \nu_\ell (p_{\nu })
\label{2.1}
\\
K^{0}(p_{1}) &\rightarrow& \pi^{-}(p_{2}) \: \ell^{+}(p_{\ell}) \: \nu_\ell(p_{\nu })
\label{2.2}
\end{eqnarray}
and their charge conjugate modes. The symbol $\ell$ stands for $e$ or $\mu $.
We work in the isospin symmetry limit ($m_{u}=m_{d})$ where all the hadronic
$K_{l3}$-decay matrix elements are equal. We therefore restrict the
discussion in the following to $K_{l3}^{0}$-decay.

In the Standard Model only the vector current $V_{\mu }=\bar{u}\gamma_{\mu
}s$ contributes to $K_{l3}$-decay, and the hadronic matrix element has
therefore the general form
\begin{equation}
\langle \pi ^{-}(p_{2})\mid \bar{u}\gamma_{\mu }s\mid K^{0}(p_{1})\rangle
=(p_{1}+p_{2})_{\mu } \: f_{+}(q^{2})
+(p_{1}-p_{2})_{\mu } \: f_{-}(q^{2})
\label{2.3}
\end{equation}
where $q=p_{1}-p_{2}$. The $q^{2}$ dependence of the form factors is usually
approximated by
\begin{equation}
f_{\pm}(q^{2})=f_{\pm}(0)\cdot \Big( 1 + \lambda_{\pm}\frac{q^{2}}{m_{\pi}^{2}}
\Big) \:.  \label{2.4}
\end{equation}

The experimental method for the determination of $\lambda_{\pm}$ consists
in comparing the measured $q^{2}$ distribution with a simulation using a
constant form factor ($\lambda_{\pm}=0$). This approximation could
possibly be too crude for future accurate data.

The slope $\lambda_{+}$ has been remeasured recently in the CPLEAR
experiment \cite{apostolakis} with the result
\begin{equation}\label{2.5}
\lambda_{+}=0.0245\pm 0.0012_{\text{\it stat}} \pm 0.0022_{\text{\it syst}} \;.
\end{equation}
This value differs by almost two standard deviations from the previous world
average \cite{caso} of
\begin{equation}
\lambda_{+}=0.0300\pm 0.0026  \label{2.7}
\end{equation}
(which is based on old data of the seventies) and from the prediction of
order $p^{4}$ chiral perturbation theory \cite{GasserLeutw2}\cite%
{LeutwylerRoos}.

The slope $\lambda_{-}$ can only be measured in $K_{\mu 3}$-decay, and its
status is even more controversial. It is common to consider also the
so-called scalar form factor (because it specifies the S-wave projection of
the crossed channel matrix element)
\begin{equation}
f_{0}(q^{2})=f_{+}(q^{2})+\frac{q^{2}}{m_{K}^{2}-m_{\pi}^{2}}f_{-}(q^{2}) \:.
\label{2.8}
\end{equation}


\section{The Lagrangian of chiral perturbation theory}

In the usual formulation of chiral perturbation theory the pseudoscalar
fields are collected in a unitary $3\times 3$ matrix
\begin{equation}
U(x)\;\;=\;\exp [i\frac{\Phi (x)}{F}]\;\;  \label{3.1}
\end{equation}
where $F$ absorbs the dimensional dependence of the fields, and, in the
chiral limit, is equal to the pion decay constant, $F=92.4$~MeV. The $3\times
3$ Matrix $\Phi $ is given by
\begin{equation}
\Phi =\lambda_{a}\phi_{a}=\left(
\begin{array}{ccc}
\pi ^{0}+\frac{1}{\sqrt{3}}\eta & \sqrt{2}\,\pi ^{+} & \sqrt{2}\,K^{+} \\
\sqrt{2}\,\pi ^{-} & -\pi ^{0}+\frac{1}{\sqrt{3}}\eta & \sqrt{2}\,K^{0} \\
\sqrt{2}\,K^{-} & \sqrt{2}\,\bar{K}^{0} & -\frac{2}{\sqrt{3}}\eta
\end{array}
\right)  \label{3.2}
\end{equation}
where $\lambda_{a}$ are the Gell-Mann matrices.

An explicit breaking of chiral symmetry is introduced via the mass matrix
\begin{equation}
\chi \;=\left(
\begin{array}{ccc}
m_{\pi}^{2} & 0 & 0 \\
0 & m_{\pi}^{2} & 0 \\
0 & 0 & 2m_{K}^{2}-m_{\pi}^{2}
\end{array}
\right)  \label{3.3}
\end{equation}
where $m_{\pi}$ and $m_{K}$ are the unrenormalized masses of the $\pi $ and
$K$-mesons. The mass of the $\eta $-meson is given to this order by the
Gell-Mann-Okubo relation
\begin{equation}
m_{\eta }^{2}=\frac{4}{3}m_{K}^{2}-\frac{1}{3}m_{\pi}^{2} \:. \label{3.4}
\end{equation}
The mass term is related to the quark masses by $\chi = \text{\it const}\cdot
\text{diag}(m_u,m_{d},m_{s})$ with $m_{u}=m_{d}$. To calculate form factors, we have
to include the interaction with external boson fields. This is done by
introducing gauge fields $l_{\mu }$, $r_{\mu }$%
\begin{equation}
l_{\mu }\;\;=\;\;\sum_{a=1}^{8} T_a l_{\mu }^{a}\qquad %
\mbox{and}\qquad r_{\mu }\;\;=\;\;\sum_{a=1}^{8} T_a r_{\mu }^{a} \;.
\label{3.6}
\end{equation}
($T_a = \lambda_{a}/2$) with their field tensors
\begin{eqnarray}
L_{\mu \nu } &=&\partial_{\mu }l_{\nu }-\partial_{\nu }l_{\mu }-i[l_{\mu
},l_{\nu }]  \label{3.7} \\
R_{\mu \nu } &=&\partial_{\mu }r_{\nu }-\partial_{\nu }r_{\mu }-i[r_{\mu
},r_{\nu }]  \label{3.8}
\end{eqnarray}
and replacing the usual derivative by a covariant one,
\begin{equation}
\partial_{\mu }\!\!\;U\;\rightarrow \;D_{\mu }\!\!\;U=\partial_{\mu
}\!\!\;U+iUl_{\mu }-ir_{\mu }U \:. \label{3.5}
\end{equation}
In this way we have extended the global chiral $SU(3)\times SU(3)$ to a
local symmetry. In case of the weak interaction, the external boson is
the $W$ and is given by
\begin{eqnarray}
l_\mu &=& -\frac{e}{\sqrt{2}\,\sin\theta_W} (W_\mu^+ T + \mbox{h.c.})
\\
r_\mu &=& 0
\end{eqnarray}
with
\begin{equation}
T \;=\; \left( \begin{array}{ccc} 0 & V_{ud} & V_{us} \\ 0&0&0 \\ 0&0&0
\end{array} \right).
\end{equation}

With the building blocks $D_{\mu }$, $U$, $U^{\dagger }$, $L_{\mu \nu }$,
$R_{\mu\nu}$ and $\chi $ we can construct the Lagrangian of chiral
perturbation theory
\begin{equation}
\mathcal{L}=\mathcal{L}^{(2)}+\mathcal{L}^{(4)}+\mathcal{L}^{(6)}+\ldots
\;\;,  \label{3.9}
\end{equation}
where $\mathcal{L}^{(2n)}$ denotes the most general expression with $2n$
powers of mass or covariant derivatives that is consistent with the
symmetries of QCD. For the two lowest orders the result is
\cite{Weinberg79}\cite{GasserLeutw1}
\begin{eqnarray}
\mathcal{L}^{(2)} &=&\frac{F^{2}}{4}\;\Tr(D_{\mu }\!\!\;UD^{\mu
}\!\!\;U^{\dagger})+\frac{F^{2}}{4}\;\Tr(\chi U^{\dagger }+U\chi ^{\dagger })
\label{3.9a} \\
\mathcal{L}^{(4)} &=&L_{1}\;\{\Tr(D_{\mu }\!\!\;UD^{\mu }\!\!\;U^{\dagger
})\}^{2}+L_{2}\;\Tr(D_{\mu }\!\!\;UD_{\nu }\!\!\;U^{\dagger })\;\Tr(D^{\mu
}\!\!\;UD^{\nu }\!\!\;U^{\dagger })  \label{3.9b} \\
&&+L_{3}\;\Tr(D_{\mu }\!\!\;UD^{\mu }\!\!\;U^{\dagger }D_{\nu }\!\!\;UD^{\nu
}\!\!\;U^{\dagger })+L_{4}\;\Tr(D_{\mu }\!\!\;UD^{\mu }\!\!\;U^{\dagger
})\;\Tr(\chi U^{\dagger }+U\chi ^{\dagger })  \notag \\
&&+L_{5}\;\Tr(D_{\mu }\!\!\;UD^{\mu }\!\!\;U^{\dagger }(\chi U^{\dagger
}+U\chi ^{\dagger }))+L_{6}\;\{\Tr(\chi U^{\dagger }+U\chi ^{\dagger })\}^{2}
\notag \\
&&+L_{7}\;\{\Tr(\chi ^{\dagger }U-U^{\dagger }\chi )\}^{2}+L_{8}\;\Tr(\chi
U^{\dagger }\chi U^{\dagger }+U\chi ^{\dagger }U\chi ^{\dagger })  \notag \\
&&-iL_{9}\;\Tr(L_{\mu \nu }D^{\mu }\!\!\;UD^{\nu }\!\!\;U^{\dagger }+R_{\mu
\nu }D^{\mu }\!\!\;U^{\dagger }D^{\nu }\!\!\;U)+L_{10}\;\Tr(L_{\mu \nu
}UR^{\mu \nu }U^{\dagger })\;.  \notag
\end{eqnarray}
The so-called low energy constants $L_{1},\ldots,L_{10}$ are unrenormalized
coupling constants which must be determined by comparison with experiment.
The $\mathcal{O}(p^{6})$ Lagrangian was determined in \cite{FearingScherer}. Out of
the 143 terms, we reproduce here only those which are relevant to
semileptonic $K$-decays:
\begin{eqnarray}
F^2 \mathcal{L}_{\mbox{\scriptsize\em rel}}^{(6)} &=& \hphantom{+\;}
\beta_{8}\;\Tr([D_{\mu }D_{\nu }U]_{\mbox{\scriptsize m}}\,\{[D^{\mu }U]_{%
\mbox{\scriptsize m}},[D^{\nu }\!\!\;\chi ]_{\mbox{\scriptsize p}}\})
\label{3.9c} \\
&&+\;\beta_{14}\;\Tr([D_{\mu }U]_{\mbox{\scriptsize m}%
}\,(\{[D^{\mu }\!\!\;\chi ]_{\mbox{\scriptsize p}},[\chi ]_{%
\mbox{\scriptsize m}}\}-\{[D^{\mu }\!\!\;\chi ]_{\mbox{\scriptsize m}},[\chi
]_{\mbox{\scriptsize p}}\}))  \notag \\
&&+\;\beta_{15}\,\{\Tr([D_{\mu }U]_{\mbox{\scriptsize m}%
}\;[D^{\mu }\!\!\;\chi ]_{\mbox{\scriptsize p}})\;\Tr([\chi ]_{%
\mbox{\scriptsize m}})  \notag \\
&&
\hphantom{+\;\beta_{15}\,\{}
-\;\Tr([D_{\mu }U]_{\mbox{\scriptsize m}}\;[\chi ]_{\mbox{\scriptsize p}%
})\;\Tr([D^{\mu }\!\!\;\chi ]_{\mbox{\scriptsize m}})\}  \notag \\
&&+\;\beta_{16}\,\{\Tr([D_{\mu }U]_{\mbox{\scriptsize m}%
}\;[D^{\mu }\!\!\;\chi ]_{\mbox{\scriptsize m}})\;\Tr([\chi ]_{%
\mbox{\scriptsize p}})  \notag \\
&&\hphantom{+\;\beta_{16}\,\{}
-\Tr([D_{\mu }U]_{\mbox{\scriptsize m}}\;[\chi ]_{\mbox{\scriptsize m}%
})\;\Tr([D^{\mu }\!\!\;\chi ]_{\mbox{\scriptsize p}})\}  \notag \\
&&+\;\beta_{17}\;\Tr([D_{\mu }U]_{\mbox{\scriptsize m}%
}\;[D^{\mu }U]_{\mbox{\scriptsize m}}\;[\chi ]_{\mbox{\scriptsize p}}\;[\chi
]_{\mbox{\scriptsize p}})  \notag \\
&&+\;\beta_{18}\;\Tr([D_{\mu }U]_{\mbox{\scriptsize m}%
}\;[D^{\mu }U]_{\mbox{\scriptsize m}}\;[\chi ]_{\mbox{\scriptsize p}})\;%
\mbox{\rm
Tr}([\chi ]_{\mbox{\scriptsize p}})  \notag \\
&&+\;\beta_{19}\;\Tr([D_{\mu }U]_{\mbox{\scriptsize m}%
}\;[D^{\mu }U]_{\mbox{\scriptsize m}})\;\;\Tr([\chi ]_{\mbox{\scriptsize
p}}[\chi ]_{\mbox{\scriptsize p}})  \notag \\
&&+\;\beta_{20}\;\Tr([D_{\mu }U]_{\mbox{\scriptsize m}%
}\;[\chi ]_{\mbox{\scriptsize p}})\;\Tr([D^{\mu }U]_{\mbox{\scriptsize
m}}\;[\chi ]_{\mbox{\scriptsize p}})  \notag \\
&&+\;\beta_{21}\;\Tr([D_{\mu }U]_{\mbox{\scriptsize m}%
}\;[D^{\mu }U]_{\mbox{\scriptsize m}})\;\Tr([\chi ]_{\mbox{\scriptsize p}%
})\;\Tr([\chi ]_{\mbox{\scriptsize p}})  \notag \\
&&+\;i\beta_{22}\;\Tr([D_{\mu }D^{\rho}\!\!\;U]_{%
\mbox{\scriptsize m}}\,[[D_{\rho}\!\!\;U]_{\mbox{\scriptsize m}},[D_{\nu
}\!\!\;G^{\mu \nu }]_{\mbox{\scriptsize p}}])  \notag \\
&&+\;i\beta_{23}\;\Tr([D_{\mu }D^{\rho}\!\!\;U]_{%
\mbox{\scriptsize m}}\,[[D_{\nu }\!\!\;U]_{\mbox{\scriptsize m}},[D_{\rho
}G^{\mu \nu }]_{\mbox{\scriptsize p}}])  \notag \\
&&+\;i\beta_{24}\;\Tr([D_{\mu }\!\!\;U]_{\mbox{\scriptsize m}%
}\;[D_{\nu }\!\!\;U]_{\mbox{\scriptsize m}}\,\{[\chi ]_{\mbox{\scriptsize
p}},[G^{\mu \nu }]_{\mbox{\scriptsize p}}\})  \notag \\
&&+\;i\beta_{25}\;\Tr([D_{\mu }\!\!\;U]_{\mbox{\scriptsize m}%
}\;[\chi ]_{\mbox{\scriptsize p}}\;[D_{\nu }\!\!\;U]_{\mbox{\scriptsize m}%
}\;[G^{\mu \nu }]_{\mbox{\scriptsize p}})  \notag \\
&&+\;i\beta_{26}\;\Tr([D_{\mu }\!\!\;U]_{\mbox{\scriptsize m}%
}\;[D_{\nu }\!\!\;U]_{\mbox{\scriptsize m}}\;[G^{\mu \nu }]_{%
\mbox{\scriptsize
p}})\;\Tr([\chi ]_{\mbox{\scriptsize p}})  \notag \\
&&+\;i\beta_{27}\;\Tr([D_{\mu }U]_{\mbox{\scriptsize m}%
}\,([[D_{\nu }\!\!\;\chi ]_{\mbox{\scriptsize m}},[G^{\mu \nu }]_{%
\mbox{\scriptsize p}}]-[[\chi ]_{\mbox{\scriptsize m}},[D_{\nu }\!\!\;G^{\mu
\nu }]_{\mbox{\scriptsize p}}]))\;,  \notag
\end{eqnarray}
where we used the notation of \cite{FearingScherer}
\begin{eqnarray}
G_{\mu \nu }\; &=&R_{\mu \nu }U+UL_{\mu \nu }  \label{3.14} \\
\lbrack A]_{\mbox{\scriptsize m}}\; &=&\frac{1}{2}(AU^{\dagger }-UA^{\dagger
})  \label{3.15} \\
\lbrack A]_{\mbox{\scriptsize p}}\; &=&\frac{1}{2}(AU^{\dagger }+UA^{\dagger
})\;.  \label{3.16}
\end{eqnarray}
with a slight change of notation for the couplings
\begin{eqnarray}
\beta_{i} &\doteq& F^{2}B_{i} \quad (\text{Fearing-Scherer})  \label{3.12a}
\end{eqnarray}
so as to use dimensionless quantities.

The Feynman rules can be derived by expanding $U=\exp (i\Phi /F)$
everywhere in $\mathcal{L}=\mathcal{L}^{(2)}+\mathcal{L}^{(4)}+\mathcal{L}%
^{(6)}$ and identifying the relevant vertex monomials in $\mathcal{L}$.
Before discussing the Feynman rules in detail in the next section,
we would like to make a remark on the definition of the form factors. The
currents entering in (\ref{2.3}) are defined on the quark level. The
connection to the effective theory is established by identifying these
currents with the Noether currents of the chiral symmetry,
\begin{eqnarray}
\bar{u}\gamma_{\mu }s &=& V_{\mu ,4}-iV_{\mu ,5}\;,  \label{3.17}
\end{eqnarray}
where $V_{a}^{\mu }=l_{a}^{\mu }+r_{a}^{\mu }$, $a=1,\ldots ,8$, denotes the
vector current in the effective theory.

It is obviously necessary to distinguish also graphically the vertices from $%
\mathcal{L}^{(2)}$, $\mathcal{L}^{(4)}$ and $\mathcal{L}^{(6)}$. We use the
conventions: a filled circle
\parbox{3mm}{\unitlength1mm\begin{picture}(0,0)
               \put(1.5,0){\circle*{2}} \end{picture}} stands for a vertex
from $\mathcal{L}^{(2)}$, a filled square
\parbox{3mm}{\unitlength1mm\begin{picture}(0,0)
              \put(1.5,0){\makebox(0,0){\rule[-1mm]{2mm}{2mm}}}
              \end{picture}} for a vertex from{\ }$\mathcal{L}^{(4)}$ and an
open square
\parbox{3mm}{\unitlength1mm\begin{picture}(0,0)
              \put(1.5,0){\begin{picture}(0,0)
                  \put(-1,1){\line(1,0){2}}
                  \put(-1,1){\line(0,-1){2}}
                  \put(1,1){\line(0,-1){2}}
                  \put(-1,-1){\line(1,0){2}}
                  \end{picture}}
              \end{picture}} for a vertex from $\mathcal{L}^{(6)}$.


\subsection{Pure meson vertices}

If loop corrections to form factor diagrams are considered various vertices
enter which involve only mesons. In the form factor calculation to $\mathcal{O}(p^{6})$
five vertices from $\mathcal{L}=\mathcal{L}^{(2)}+\mathcal{L}^{(4)}+\mathcal{%
L}^{(6)}$ will contribute:

From $\mathcal{L}^{(2)}$ we have to consider vertices with 4 and 6 meson fields
\begin{eqnarray}
48F^2 \, \mathcal{L}_{\text{\it four}}^{(2)} &=&
2 \Tr([\phi ,\partial_{\mu }\phi]\phi \, \partial^{\mu }\phi )
+ \Tr(\chi \phi^{4})
\label{4.1}
\end{eqnarray}
and
\begin{eqnarray}
1440 F^4 \, \mathcal{L}_{\text{\it six}}^{(2)} &=& \hphantom{+\;}
2\Tr(\partial_{\mu }\phi \: \partial^{\mu }\phi \: \phi^{4})
-8\Tr(\partial_{\mu }\phi \: \phi \partial^{\mu }\phi \: \phi^{3})
\\
&& + \; 6\Tr(\partial_{\mu }\phi \: \phi^{2} \partial^{\mu}\phi \: \phi^{2})
-\Tr(\chi \phi^{6}) \:.  \notag
\end{eqnarray}
Similarly we have the two- and four-meson vertex from $\mathcal{L}^{(4)}$%
\begin{eqnarray}
F^2 \, \mathcal{L}_{\text{\it two}}^{(4)} &=& \hphantom{+\;}
2L_{4} \: \Tr(\partial_{\mu }\phi \: \partial^{\mu }\phi) \: \Tr(\chi ) \label{4.3} \\
&&+\; 2L_{5} \: Tr\;(\chi \,\partial_{\mu }\phi \: \partial^{\mu}\phi )  \notag \\
&&-\;4L_{6} \: \Tr(\chi \phi^{2})\;\Tr(\chi )  \notag \\
&&-\; 4L_{7} \: \{\Tr(\chi \phi )\}^{2}  \notag \\
&&-\;2L_{8} \: \{\Tr(\chi ^{2}\phi^{2})+\Tr(\chi \phi \chi \phi )\},
\notag
\end{eqnarray}
\begin{eqnarray}
6F^4 \, \mathcal{L}_{\text{\it four}}^{(4)} &=&\hphantom{+\;}
6L_{1} \: \{\Tr(\partial_{\nu }\phi \: \partial^{\nu}\phi )\}^{2} \label{3.27} \\
&&+\; 6 L_{2} \: \Tr(\partial_{\mu }\phi \: \partial_{\nu}\phi )
              \; \Tr(\partial^{\mu }\phi \: \partial^{\nu }\phi ) \notag \\
&&+\; 6 L_{3} \: \Tr(\partial_{\mu }\phi \: \partial^{\mu}\phi \:
                     \partial_{\nu }\phi \: \partial^{\nu }\phi )  \notag \\
&&-\;2 L_{4} \: \{\Tr([\phi, \partial_{\nu }\phi]\phi \: \partial^{\nu }\phi )
                    \;\Tr(\chi )
                +3 \Tr(\partial_{\nu }\phi \: \partial^{\nu }\phi )
                    \;\Tr(\chi \phi^{2})\}  \notag \\
&&-\; L_{5} \: \{ 2\Tr(\chi \phi^{2}\partial_{\nu }\phi \:
                    \partial^{\nu }\phi )
                +3\Tr(\chi \phi \: \partial_{\nu }\phi \: \partial^{\nu }\phi \: \phi )
                -\Tr(\chi \phi \: \partial_{\nu}\phi \: \phi \: \partial^{\nu}\phi)  \notag \\
&&\hphantom{-\; L_{5} \: \{}
                +\; \Tr(\chi \,\partial_{\nu }\phi \: \phi^{2}
                        \partial^{\nu}\phi )
                -\Tr(\chi \,\partial_{\nu }\phi \: \phi \: \partial^{\nu }\phi \: \phi )
                +2\Tr(\chi \, \partial_{\nu }\phi \: \partial^{\nu }\phi \: \phi^{2}) \} \notag \\
&&+\; 2 L_{6} \: \{\Tr(\chi \phi^{4}) \; \Tr(\chi )+3[\Tr(\chi \phi^{2})]^{2}\}  \notag \\
&&+\; 8 L_{7} \: \Tr(\chi \phi^{3})\;\Tr(\chi \phi )  \notag \\
&&+\; L_{8} \: \{ \Tr(\chi ^{2}\phi^{4})+2\Tr(\chi \phi \chi \phi
^{3})+3\Tr(\chi \phi^{2}\chi \phi^{2})+2\Tr(\chi \phi^{3}\chi \phi )\}.
\notag
\end{eqnarray}
The parameters $L_{1},\ldots ,L_{10}$, the so called low energy constants of
$\mathcal{L}^{(4)}$, are not fixed by the symmetries but must be determined
by comparing perturbative results with experimental data or with models. The low
energy constants also serve to renormalize the loop diagrams. Therefore they
contain divergent pieces which, in dimensional regularization, manifest
themselves in $\frac{1}{\varepsilon }$-poles $(D=4-2\varepsilon )$ and which
were calculated in \cite{GasserLeutw1}.

Finally, the two-meson vertex from $\mathcal{L}^{(6)}$ enters:
\begin{eqnarray}
F^4 \mathcal{L}_{\text{\it two}}^{(6)} &=&
-\beta_{17}\;\Tr(\partial_{\mu}\phi \: \partial^{\mu }\phi \,\chi \chi )  \label{4.4} \\
&&-\beta_{18}\;\Tr(\partial_{\mu }\phi \: \partial^{\mu}\phi \,\chi )
              \;\Tr(\chi)  \notag \\
&&-\beta_{19}\;\Tr(\partial_{\mu}\phi \: \partial^{\mu}\phi )
              \;\Tr(\chi \chi )  \notag \\
&&-\beta_{20}\;\Tr(\partial_{\mu }\phi \,\chi)
              \;\Tr(\partial^{\mu }\phi \,\chi )  \notag \\
&&-\beta_{21}\;\Tr(\partial_{\mu }\phi \: \partial^{\mu}\phi )
              \;\Tr(\chi )\;\Tr(\chi ) \:. \notag
\end{eqnarray}


\subsection{W-boson-meson vertices}

Every diagram contributing to $K_{l3}$-decay contains one vertex where the
external $W$-boson couples to the mesons. The Feynman rules of the
corresponding vertices result from the terms in $\mathcal{L}$ that are
linear in the gauge fields. Thus, the left-handed and right-handed mesonic
currents that couple to the external pseudo-scalar mesons are given by
\begin{equation}
J_{\mu ,a}^{L}\;\;\;=\;\;\frac{\delta \mathcal{L}}{\delta l^{\mu ,a}}
\bigg|_{r_{\mu }=l_{\mu }=0}
\; , \quad
J_{\mu ,a}^{R}\;\;\;=\;\;\frac{\delta \mathcal{L}}{\delta r^{\mu ,a}}
\bigg|_{r_{\mu }=l_{\mu }=0} \:.
\label{3.18}
\end{equation}
The result for the currents from $\mathcal{L}^{(2)}$ reads
\begin{eqnarray}
J_{\mu ,a}^{L}[\mathcal{L}^{(2)}] &=& \hphantom{+\;}
\frac{i}{4}\;\Tr(T_{a}[\partial_{\mu}\phi ,\phi ])  \label{3.18a} \\
&&
-\; \frac{i}{48F^{2}}\;\mbox{\rm Tr}\,T_{a}([\partial_{\mu }\phi ,\phi
^{3}]-3\phi [\partial_{\mu }\phi ,\phi ]\phi )  \notag \\
&&
+\;\frac{i}{1440F^{4}}\;\mbox{\rm Tr}\,T_{a}([\partial_{\mu }\phi ,\phi
^{5}]-5\phi \lbrack \partial_{\mu }\phi ,\phi^{3}]\phi +10\phi
^{2}[\partial_{\mu }\phi ,\phi ]\phi^{2})  \notag
\end{eqnarray}
where we have only written the terms contributing to $K_{l3}$-decay at
$\mathcal{O}(p^6)$. We
represent a $W$-vertex from $\mathcal{L}^{(2)}$ by a filled circle
\parbox{3mm}{\unitlength1mm\begin{picture}(0,0)%
               \put(1.5,0){\circle*{2}} \end{picture}}.

The result for the relevant terms of the current from $\mathcal{L}^{(4)}$
reads
\begin{eqnarray}
F^2 \: J_{\mu,a}^{L}[\mathcal{L}^{(4)}]_{\text{\it two}}
&=& \hphantom{+\;}
2iL_{4} \: \Tr(T_{a}[\partial_{\mu }\phi ,\phi]) \; \Tr(\chi )
\\
&&
+\; iL_{5} \: \mbox{\rm Tr}\,T_{a}(\chi
\,\partial_{\mu}\phi \: \phi + \partial_{\mu }\phi \: \chi \phi
-\phi \chi \partial_{\mu }\phi - \phi \,\partial_{\mu }\phi \: \chi )
\notag \\
&&-\; iL_{9} \: \partial^{\nu}\,\Tr(T_{a}[\partial_{\mu}\phi,
\partial_{\nu }\phi])  \notag
\end{eqnarray}
for the two-meson-$W$ vertex and
\begin{eqnarray}
12F^4 \: J_{\mu ,a}^{L}[\mathcal{L}^{(4)}]_{\text{\it four}} &=&
\hphantom{+\;}
24iL_{1} \: \Tr(T_{a}[\partial_{\mu }\phi,\phi ]) \;
\Tr(\partial_{\nu }\phi \; \partial^{\nu }\phi)  \label{3.21} \\
&&
+\; 24iL_{2} \: \Tr(T_{a}[\partial^{\nu }\phi, \phi]) \;
                \Tr(\partial_{\mu }\phi \; \partial_{\nu }\phi)  \notag \\
&&
+\; 12 i L_{3} \: \mbox{\rm Tr}\,T_{a}(
\{\partial_{\mu }\phi, \partial_{\nu }\phi \;
\partial^{\nu }\phi \} \phi - \phi \{
\partial_{\mu }\phi, \partial_{\nu }\phi \:
\partial^{\nu }\phi \})  \notag \\
&&
-\; 2 i L_{4} \{ 6\Tr(T_{a}[\partial_{\mu }\phi,\phi])
\Tr(\chi\phi^{2})+\Tr T_{a}([\partial_{\mu }\phi,\phi^{3}]
-3\,\phi [\partial_{\mu }\phi,\phi]\phi) \; \Tr(\chi ) \}  \notag \\
&&
-\; i L_{5} \: \{\mbox{\rm Tr}\,T_{a}(\chi
[ \partial_{\mu}\phi ,\phi ] \phi^{2}
+2\,\chi \phi^{2} \partial_{\mu }\phi \: \phi
-2\,\phi \chi \{\partial_{\mu }\phi, \phi^{2}\}
) \notag \\
&&
\qquad\quad
+ \mbox{\rm Tr}\,T_{a}(
2\,\partial_{\mu }\phi \: \chi \phi^{3}
+4\,\phi \chi \phi \,\partial_{\mu }\phi \: \phi
+3\,[\partial_{\mu }\phi, \phi]\chi \phi^{2}
)  \notag \\
&&
\qquad\quad
+\mbox{\rm Tr}\,T_{a}(
3\,\phi^{2}\chi [\partial_{\mu}\phi,\phi]
+2\,\{\partial_{\mu }\phi ,\phi^{2}\}\chi\phi
-4\,\phi \,\partial_{\mu }\phi \: \phi \chi\phi
)  \notag \\
&&
\qquad\quad
+\mbox{\rm Tr}\,T_{a}(
-2\,\phi^{3}\chi \, \partial_{\mu }\phi
-2\,\phi \,\partial_{\mu }\phi \: \phi^{2}\chi
+\phi^{2}[\partial_{\mu }\phi,\phi]\chi )\}  \notag \\
&&
+\; i L_{9} \partial^{\nu} \{ \mbox{\rm Tr}\,T_{a}(
-3\,\phi [ \partial_{\mu }\phi, \partial_{\nu }\phi]\phi
+\partial_{\mu }\phi \: \phi^{2} \, \partial_{\nu }\phi
-\partial_{\nu}\phi \: \phi^{2}\,\partial_{\mu }\phi )  \notag \\
&&
\qquad\quad
+\mbox{\rm Tr}\,T_{a}(
2\,\phi^{2}[\partial_{\mu }\phi, \partial_{\nu }\phi]
+2\,[\partial_{\mu }\phi, \partial_{\nu}\phi ] \phi^{2}
- [\partial_{\mu }\phi \: \phi,\partial_{\nu }\phi \; \phi]
\notag \\
&&
\qquad\quad
-\mbox{\rm Tr}\,T_{a}(
[\phi \,\partial_{\mu }\phi, \phi \,\partial_{\nu }\phi ])\}
\notag
\end{eqnarray}
for the four-meson vertex with one $W$-boson.
In the diagrams we represent a $W$-vertex from $\mathcal{L}^{(4)}$
by a filled square
\parbox{3mm}{\unitlength1mm\begin{picture}(0,0)
              \put(1.5,0){\makebox(0,0){\rule[-1mm]{2mm}{2mm}}}
              \end{picture}}.

Finally a 2-meson-$W$ vertex from $\mathcal{L}^{(6)}$ contributes:
\begin{eqnarray}
4F^4 \, J_{\mu ,a}^{L}[\mathcal{L}^{(6)}] &=& \hphantom{+\;}
2 i \beta_{8} \: \Tr\,T_{a}[\chi, \{\partial_{\mu }\partial_{\nu }\phi,
                         \partial^{\nu }\phi \}]  \label{3.22} \\
&&
+\; i \beta_{14} \: \Tr\,T_{a}(2\!\ \chi \!\ \phi \, \partial_{\mu }\phi \: \chi
                         - 2\!\ \chi \, \partial_{\mu }\phi \: \phi \, \chi
                         + \chi \chi \, \partial_{\mu }\phi \: \phi  \notag \\
&&\hphantom{+\; i \beta_{14} \: \Tr\,T_{a}(}
-\;\chi \chi \!\;\phi \,\partial_{\mu }\phi
+\partial_{\mu }\phi \: \phi \!\; \chi \chi
-\phi \,\partial_{\mu }\phi \: \chi \chi )  \notag \\
&&
+\; 2 i \beta_{15} \: \{ \Tr(\chi \phi ) \;
\Tr\,T_{a}[\partial_{\mu }\phi, \chi ]
- \Tr(\chi \,\partial_{\mu }\phi) \; \Tr\,T_{a}[\phi ,\chi ]\}  \notag \\
&&
+\; i \beta_{16} \: \Tr(\chi ) \; \Tr\,T_{a}[\{\partial_{\mu }\phi ,\phi \},\chi ]  \notag \\
&&
+\; 2 i \beta_{17} \: \Tr\,T_{a}(\phi \,\partial_{\mu}\phi \: \chi \chi
-\partial_{\mu }\phi \: \chi \chi \phi
+\phi \chi\chi \,\partial_{\mu }\phi
-\chi \chi \,\partial_{\mu }\phi \,\phi )  \notag \\
&&
+\; 2 i \beta_{18} \;\; \Tr(\chi )\;\Tr\,T_{a}(
\phi\,\partial_{\mu }\phi \: \chi
-\partial_{\mu }\phi \: \chi \,\phi
+\phi \, \chi \, \partial_{\mu }\phi
-\chi \, \partial_{\mu }\phi \, \phi )  \notag \\
&&
+\;4 i \beta_{19}\,\{\Tr(\chi \chi ) \;
                     \Tr\,(T_{a}[\phi, \partial_{\mu }\phi])
      +2\,\Tr(\chi \lbrack \chi ,\phi]) \; \Tr(T_{a}\,\partial_{\mu }\phi )\}  \notag \\
&&
+\; 4 i \beta_{20} \: \Tr(\chi \partial_{\mu}\phi) \; \Tr\,(T_{a}[\phi,\chi])  \notag \\
&&
+\; 4 i \beta_{21} \: \Tr(\chi )\;\Tr(\chi )\;\Tr(T_{a}[\phi,\partial_{\mu }\phi ])  \notag \\
&&
+\; 4 i \beta_{22} \: \Tr\,T_{a}(\partial_{\nu}\partial^{\nu }
           [\partial_{\mu}\partial_{\rho}\phi, \partial^{\rho}\phi]
         -\partial_{\mu}\partial^{\nu}[\partial_{\nu}\partial_{\rho}\phi,
         \partial^{\rho}\phi])  \notag \\
&&
+\; 4 i \beta_{23} \: \partial^{\nu}\partial^{\rho}\;\! \Tr\,T_{a}(
         [\partial_{\mu}\partial_{\rho}\phi, \partial_{\nu}\phi]
        -[\partial_{\nu}\partial_{\rho}\phi, \partial_{\mu}\phi]) \notag \\
&&
-\; 4 i \beta_{24} \: \partial^{\nu }\;\!\Tr(
         [\partial_{\mu}\phi, \partial_{\nu }\phi] \{\chi ,T_{a}\})  \notag \\
&&
- 4 i \beta_{25} \: \partial^{\nu }\;\!\Tr\,T_{a}(
         \partial_{\mu}\phi \: \chi \;\!\partial_{\nu}\phi
        -\partial_{\nu }\phi \: \chi \;\!\partial_{\mu}\phi )  \notag \\
&&
-\; 4 i \beta_{26} \: \partial^{\nu}\;\!\Tr(T_{a}
         [\partial_{\mu}\phi, \partial_{\nu }\phi])
         \; \Tr(\chi )  \notag \\
&&
+\; 2 i \beta_{27} \: \partial^{\nu }\;\!\Tr\,T_{a}(
         [\partial_{\mu}\phi, \{\partial_{\nu }\phi, \chi \}]
        -[\partial_{\nu}\phi, \{\partial_{\mu }\phi ,\chi \}]) \:. \notag
\end{eqnarray}
A $W$-vertex from $\mathcal{L}^{(6)}$ is represented by an open square
\parbox{3mm}{\unitlength1mm\begin{picture}(0,0)
              \put(1.5,0){\begin{picture}(0,0)
                  \put(-1,1){\line(1,0){2}}
                  \put(-1,1){\line(0,-1){2}}
                  \put(1,1){\line(0,-1){2}}
                  \put(-1,-1){\line(1,0){2}}
                  \end{picture}}
              \end{picture}}.
From the interactions given above, the Feynman rules can be extracted by
transforming into momentum space and symmetrizing over the meson fields.


\subsection{Renormalization scheme}

Before evaluating the loop diagrams, we must specify the regularization and
renormalization scheme. In our calculation we are using dimensional
regularization and the so called \emph{GL}-scheme which is defined in the
following way: each diagram of order $\mathcal{O}(p^{2n})$ is multiplied
with a factor $e^{(1-n)\alpha (\varepsilon )}$ where $D=4-2\varepsilon $ is
the dimension of space-time and $\alpha (\varepsilon )$ is given by
\begin{equation}
(4\pi )^{\varepsilon }\,\Gamma (-1+\varepsilon )=-\frac{e^{\alpha
(\varepsilon )}}{\varepsilon }\;,  \label{5.40}
\end{equation}
that is
\begin{eqnarray}
\alpha (\varepsilon ) &=& \varepsilon (1-\gamma +\log 4\pi )+\varepsilon ^{2}(%
\frac{\pi ^{2}}{12}+\frac{1}{2})+\mathcal{O}(\varepsilon ^{3})\,\;.\quad
\label{5.41}
\end{eqnarray}
Because of $\alpha (0)=0$ the total $\mathcal{O}(p^{6})$ result is unchanged
in $D=4$ dimensions. The reason for this modification of each diagram is to
eliminate the geometric factor $(4\pi )^{\varepsilon }\,\Gamma
(-1+\varepsilon )$ appearing in the 1-loop integrals. This renormalization
scheme is very similar to the well-known $\overline{\text{MS}}$ scheme,
where each diagram is multiplied by a factor $(4\pi )^{-\varepsilon
}\,e^{\gamma \varepsilon }$ per loop (instead of $e^{-\alpha (\varepsilon )}$%
).

The GL-scheme extends the usual 1-loop scheme introduced by Gasser and
Leutwyler \cite{GasserLeutw1} in a natural way. This can be understood by
considering the renormalization constants $L_{i}$ of $\mathcal{L}^{(4)}$: In
$D$-dimensional space-time they have dimension $D-4$ and their dimension is
made manifest by the mass scale $\mu $ of dimensional regularization:
\begin{equation}
L_{i}=\mu ^{D-4}\,L_{i}(\mu ,D)\;.  \label{5.41a}
\end{equation}
$L_{i}(\mu ,D)$ has the same $\mu $-dependence as a 1-loop integral, because
$L_{i}$ itself is independent of $\mu $. It can be expanded in a Laurent
series around $\varepsilon =0$ in the same way as a 1-loop integral:
\begin{equation}
L_{i}(\mu ,D) =
\frac{L_{i}^{(-1)}}{\varepsilon }
+ L_{i}^{(0)}(\mu )
+ \varepsilon \, L_{i}^{(1)}(\mu )
+\mathcal{O}(\varepsilon^2)\;.  \label{5.41b}
\end{equation}
In the usual 1-loop scheme one chooses
\begin{eqnarray}
L_{i}^{(-1)} &=&-\frac{\Gamma_{i}}{32\pi ^{2}}  \label{5.42} \\
L_{i}^{(0)}(\mu ) &=&L_{i}^{\text{\it ren}}(\mu )-\frac{\Gamma_{i}}{32\pi ^{2}}%
[1-\gamma +\log 4\pi] \:,  \label{5.43}
\end{eqnarray}
where $\Gamma_{i}$ are numbers which can be found in \cite{GasserLeutw1}.
The second term in $L_{i}^{(0)}$ is constructed so that it cancels in the $%
\varepsilon ^{0}$-coefficient after multiplication with the \emph{GL}-factor
$e^{-\alpha(\varepsilon )}$:
\begin{equation}
\label{5.44}
L_{i}^{GL}(\mu ,D) \;=\;
e^{-\alpha (\varepsilon )}\!L_{i}(\mu ,D)
\;=\; \frac{L_{i}^{(-1)}}{\varepsilon }
+ L_{i}^{(0),GL}(\mu )
+ \varepsilon \, L_{i}^{(1),GL}(\mu )
+\mathcal{O}(\varepsilon^2)
\end{equation}
with $L_{i}^{(0),GL}(\mu )=L_{i}^{\text{\it ren}}(\mu )$.

The dimension of the $\mathcal{L}^{(6)}$-parameters $\beta_{i}$ appearing
in (\ref{4.4}) can be treated in the same way:
\begin{equation}
\beta_{i}=\mu ^{2D-8} \: \beta_{i}(\mu ,D) \:,  \label{5.45}
\end{equation}
where $\beta_{i}(\mu ,D)$ behaves like a 2-loop integral. Its Laurent
series in the above $GL$-scheme is given by
\begin{equation}
\beta_{i}^{GL}(\mu ,D) \;=\;
e^{-2\alpha (\varepsilon )}\!\beta_{i}(\mu ,D) \;=\;
\frac{\beta_{i}^{(-2)}}{\varepsilon ^{2}}+\frac{\beta_{i}^{(-1),GL}(\mu
)}{\varepsilon }+\beta_{i}^{(0),GL}(\mu )+\mathcal{O}(\varepsilon )\;.
\label{betaGL}
\end{equation}


\subsection{Mass- and wavefunction renormalization}

To order $p^{6}$, finite S-matrix elements in chiral perturbation theory are
obtained by multiplying the unrenormalized one-particle irreducible (1PI)
Feynman diagrams obtained from $\mathcal{L}=\mathcal{L}^{(2)}+\mathcal{L}%
^{(4)}+\mathcal{L}^{(6)}$ with a factor $\sqrt{Z}$ per external meson, where $Z$ is the wave
function renormalization constant for this meson.
To be on familiar ground, we
start by calculating the mass- and wave function renormalization
from the renormalized propagator
\begin{equation}
\frac{i}{p^{2}-m^{2}-\Sigma (p^{2})},  \label{5.1}
\end{equation}
where $m$ denotes the bare meson mass and $\Sigma(p^{2})$ the 1PI
unrenormalized self energy which is given perturbatively by
\begin{equation}
\Sigma (p^{2})=\Sigma_{1}(p^{2})+\Sigma_{2}(p^{2})+\ldots \:.  \label{5.2}
\end{equation}
The leading $\mathcal{O}(p^{2})$-contribution vanishes, so that $\Sigma_{1}$%
, $\Sigma_{2}$ represent the contributions of $\mathcal{O}(p^{4})$ and $%
\mathcal{O}(p^{6})$.

From the condition that the renormalized propagator develops a pole with
residue 1 at $p^{2}=m_{ph}^{2}$, where $m_{ph}$ is the physical or pole
mass, one derives the conditions
\begin{equation}
\delta m^{2} = m_{ph}^{2}-m^{2} = \Sigma (p^{2}=m_{ph}^{2},m^{2},F)  \label{5.3}
\end{equation}
\begin{equation}
Z^{-1}=1-\Sigma ^{\prime }(p^{2}=m_{ph}^{2},m^{2},F)  \label{5.4}
\end{equation}
where $m$ stands symbolically for the set of unrenormalized masses and $F$
is the unrenormalized pion decay constant (all assumed to be given as
functions of the renormalized or physical parameters). Perturbatively, $Z$ is
therefore given by
\begin{equation}
Z=1+\delta Z_{1}+\delta Z_{2}+\ldots  \label{5.5}
\end{equation}
with the $\mathcal{O}(p^{4})$- and $\mathcal{O}(p^{6})$-corrections
\begin{eqnarray}
\delta Z_{1} &=&\Sigma_{1}^{\prime }(p^2=m^2,m^{2},F)  \label{5.6} \\
\delta Z_{2} &=&\Sigma_{2}^{\prime }(m^{2}_{\text{\it ph}})
+\Sigma_{1}^{\prime}(m^{2}_{\text{\it ph}})^{2}.  \label{5.7}
\end{eqnarray}
where we have expanded $\Sigma_1^\prime$ around $p^2=m^2$ and used
the fact that the term involving the second derivative of
$\Sigma_{1}(p^{2})$ vanishes,
i.e. $\Sigma_1^\prime$ is independent of $p^2$.
In eq. (\ref{5.6}) the unrenormalized quantities $m^{2},F$ must be
expressed by their physical values according to eq.~(\ref{5.3}) and
eq.~(\ref{5.7e}) below.

Each external meson propagator must be multiplied with a factor%
\begin{equation}
\sqrt{Z}\;=\;\sqrt{1+\delta Z}\;=\;1+\frac{\delta Z_{1}}{2}+[\frac{\delta
Z_{2}}{2}-\frac{(\delta Z_{1})^{2}}{8}]+\mathcal{O}(p^{6}) \:. \label{5.70}
\end{equation}

We have to calculate $\Sigma (p^{2})$ for $\pi$-, $K$-mesons in order to
determine $Z_{\pi },Z_{K}$ and $\delta m_{\pi}^2$, $\delta m_{K}^2$. The $%
\mathcal{O}(p^{4})$ results are well known \cite{GasserLeutw1}:
\begin{eqnarray}
\delta Z_1^\pi &=&
-\frac{1}{3F^2} \{
A_0(m_K^2) + 2 A_0(m_\pi^2)
\label{Zpi_oneloop}
\\
&&\hphantom{-\frac{1}{3F^2} \{}
+ 24 L_4 (2 m_K^2 + m_\pi^2) + 24 L_5 m_\pi^2
\} \notag
\\
\delta Z_1^K &=&
-\frac{1}{4F^2} \{
A_0(m_\eta^2)+2 A_0(m_K^2)+A_0(m_\pi^2)
\\
&&\hphantom{-\frac{1}{4F^2} \{}
+ 32 L_4 (2 m_K^2 + m_\pi^2) + 32 L_5 m_K^2
\} \notag
\\
\delta m_\pi^2 &=&
\frac{1}{6 F^2} \: \{
m_\pi^2 \: A_0(m_\eta^2)
-3 m_\pi^2 \: A_0(m_\pi^2)
\\&&\hphantom{\frac{1}{6 F^2} \: \{}
-48 m_\pi^2 (2 m_K^2+m_\pi^2) \: L_4
-48 L_5 m_\pi^4 \notag
\\&&\hphantom{\frac{1}{6 F^2} \: \{}
+96 L_6 m_\pi^2 (2 m_K^2 + m_\pi^2)
+96 L_8 m_\pi^4
\}\notag
\\
\delta m_K^2 &=&
\frac{1}{12 F^2} \: \{
- 4 m_K^2 \: A_0(m_\eta^2)
-96 L_4 m_K^2 (2 m_K^2+m_\pi^2)
\\&&\hphantom{\frac{1}{12 F^2} \: \{}
-96 L_5 m_K^4
+192 L_6 m_K^2 (2 m_K^2 + m_\pi^2)
+192 L_8 m_K^4
\}\notag
\end{eqnarray}
The function $A_{0}(m^{2})$ is the standard tadpole integral
\begin{equation}
A_{0}(m^{2})=\mu ^{4-D}\int \frac{d^{D}k}{i(2\pi )^{D}}\frac{1}{k^{2}-m^{2}}%
\;,  \label{5.7a1}
\end{equation}
where $D=4-2\varepsilon$ is the dimension of space-time.

In addition, we need the renormalization of the pion decay constant%
\begin{equation}
\delta F=F_{\pi }-F \:, \label{5.7e}
\end{equation}
where $F_{\pi }$ is the physical pion decay constant. We only quote the
result \cite{GasserLeutw1}:
\begin{equation}
\delta F = \frac{1}{2F}\{A_{0}(m_{K}^{2})+2A_{0}(m_{\pi
}^{2})+8L_{4}(2m_{K}^{2}+m_{\pi}^{2})+8L_{5}m_{\pi}^{2}\} \:. \label{5.7f}
\end{equation}

It should be noted, that the renormalization constants $\delta m^{2}$ and $%
\delta F$ defined above are finite.
The divergences and scale dependence of the loop integrals are canceled by
similar factors in the counter terms $L_{i}$ from $\mathcal{L}^{(4)}$.

The self energy diagrams contributing up to two loops and to order $p^{6}$
are given in fig.~1. External legs are fixed by the mesons
considered, while one has to sum over all possible internal meson lines.

We call a two-loop diagram ''reducible'', if the two loop integrations
decouple, i.e. if they are given by a product of one-loop integrals.
Otherwise they are called ''irreducible''.
The $\mathcal{O}(p^6)$-correction of the wave function renormalization
$Z$ consists of three parts
\begin{equation}
  \delta Z_2 = \delta Z_2^{\textit{red}} + \delta Z_2^{\textit{irred}}
               + \delta Z_2[\mathcal{L}^{(6)}]
\end{equation}
which are given below for $\pi$ and $K$.

The one-loop and the reducible two-loop diagrams of fig.~1 yield for the pion
wave function renormalization $Z^{\pi }$ an $\mathcal{O}(p^6)$-contribution%
\begin{eqnarray}
\delta Z_2^{\pi,\textit{red}} &=&
\frac{1}{180 F^4} \{
\hphantom{+}
2880 \: \big[ 2 L_1 + 4 L_2 + L_3 \big] \: A_2(m_\pi^2)
\\ \nonumber && \hphantom{\frac{1}{180 F^4} \Big\{}
+ 2880 \: \big[ 4 L_2 + L_3 \big] \: A_2(m_K^2)
+ 960 \: \big[ 3 L_2 + L_3 \big] \: A_2(m_\eta^2)
\\ \nonumber && \hphantom{\frac{1}{180 F^4} \Big\{}
+ 960 m_\pi^2 \: \big[ (2 m_K^2 + m_\pi^2) (L_4 - 2 L_6) + m_\pi^2 (L_5 - 2 L_8) \big] \: A_{0;2}(m_\pi^2)
\\ \nonumber && \hphantom{\frac{1}{180 F^4} \Big\{}
+ 480 m_K^2 \: \big[ (m_\pi^2 + 2 m_K^2) (L_4 - 2 L_6) + m_K^2 (L_5 - 2 L_8) \big] \: A_{0;2}(m_K^2)
\\ \nonumber && \hphantom{\frac{1}{180 F^4} \Big\{}
+ 720 m_\pi^2 \: \big[ 12 L_1 + 4 L_2 + 6 L_3 - 6 L_4 - 3 L_5 \big] \: A_0(m_\pi^2)
\\ \nonumber && \hphantom{\frac{1}{180 F^4} \Big\{}
+ 480 \: \big[ 24 m_K^2 L_1 + 6 m_K^2 L_3 - 12 m_K^2 L_4 - (m_\pi^2 + 2 m_K^2) L_5 \big] \: A_0(m_K^2)
\\ \nonumber && \hphantom{\frac{1}{180 F^4} \Big\{}
+ 80 \: \big[ 2(4 m_K^2 - m_\pi^2)(6 L_1 + L_3 - 3 L_4) - 3 m_\pi^2 L_5 \big] \: A_0(m_\eta^2)
\\ \nonumber && \hphantom{\frac{1}{180 F^4} \Big\{}
+ 20 m_\pi^2 \: \big[ 3 A_0(m_\pi^2) - A_0(m_\eta^2) \big] \: A_{0;2}(m_\pi^2)
+ 20 m_K^2 \: A_0(m_\eta^2) \: A_{0;2}(m_K^2)
\\ \nonumber && \hphantom{\frac{1}{180 F^4} \Big\{}
- 15 \: A_0(m_K^2) \: A_0(m_\pi^2)
+ 6 \: A_0(m_K^2)^2
+ 9 \: A_0(m_\eta^2) \: A_0(m_K^2)
\}
\\ \nonumber &&
+\; ( \delta Z_1^{\text{$\pi$}} )^2 \:,
\end{eqnarray}
where $\delta Z_1^\pi$ is the $\mathcal{O}(p^4)$ result from
eq.~(\ref{Zpi_oneloop}) and the additional functions $A_{0;2}$ and $A_2$
are related to the tadpole integral $A_0$ and the dimension $D$
of space-time (cf. eqs.~(\ref{A02_deriv_definition}) and (\ref{A2_definition})
in appendix \ref{OneLoopIntegrals}):
\begin{equation}\label{A0_and_A2}
  A_{0;2}(m^2) 
                  = \frac{1}{m^2} \Big(\frac{D}{2} - 1 \Big) \: A_0(m^2)
  \quad\text{and}\quad
  A_2(m^2) = \frac{m^2}{D} \: A_0(m^2) \:.
\end{equation}

For the irreducible two-loop contributions of fig.~1, which involve
higher transcendental functions, we only quote the exact result for the
divergent part and a numerical result for the finite part. The latter
involves an arbitrary scale $\mu $ which cancels in the final answer for the
form factor. {For the choice}
\begin{equation}
\label{massenwerte}
\mu=m_\rho=770\;\text{MeV}, \; m_K=495\;\text{MeV},
\; m_{\pi}=0.28\:m_{K}, \; F_\pi = 92.4 \; \text{MeV,}
\end{equation}
$m_\eta^2$ being an abbreviation for the Gell-Mann-Okubo term
$\frac{4}{3}m_K^2 - \frac{1}{3}m_\pi^2$, and the definition
\begin{equation*}
\mathrm{Lg}(m^{2}) \;\doteq\; \log (\frac{m^{2}}{4\pi \mu ^{2}})+\gamma
\end{equation*}%
we obtain%
\begin{eqnarray}
\delta Z_2^{\pi,\textit{irred}} &=&
2.015180506  
+ \frac{78 m_K^4+22 m_K^2 m_\pi^2+89 m_\pi^4}
       {72(4\pi)^4 F^4 \, \varepsilon^2}
\\&&
+\; \frac{1}{288(4\pi)^4 F^4 \, \varepsilon} \Big\{
1452 m_K^4
-36 m_K^2 m_\pi^2
+1163 m_\pi^4
\notag
\\&&\qquad\qquad
-4 (24 m_K^4+14 m_K^2 m_\pi^2-5 m_\pi^4) \: \mathrm{Lg}(m_\eta^2)
\nonumber\\&&\qquad\qquad
-528  m_K^4 \: \mathrm{Lg}(m_K^2)
\nonumber\\&&\qquad\qquad
-12  m_\pi^2 (10 m_K^2+61 m_\pi^2) \: \mathrm{Lg}(m_\pi^2)
\Big\} \:. \nonumber
\end{eqnarray}

Similarly we obtain for the reducible part of the kaon wave function
renormalization
\begin{eqnarray}
\delta Z_2^{\text{$K,\textit{red}$}} &=&
\frac{1}{360 F^4} \{
4320 \: \big[ 4 L_2 + L_3 \big] \: A_2(m_\pi^2)
\\ \nonumber && \hphantom{\frac{1}{360 F^4} \{}
+2880 \: \big[ 4 L_1 + 10 L_2 + 3 L_3 \big] \: A_2(m_K^2)
+480 \: \big[ 12 L_2 + L_3 \big] \: A_2(m_\eta^2)
\\ \nonumber && \hphantom{\frac{1}{360 F^4} \{}
+720 m_\pi^2 \: \big[ (2 m_K^2 + m_\pi^2) (L_4 - 2 L_6)
             + m_\pi^2 (L_5 - 2 L_8) \big] \: A_{0;2}(m_\pi^2)
\\ \nonumber && \hphantom{\frac{1}{360 F^4} \{}
+1440 m_K^2 \: \big[ (2 m_K^2 + m_\pi^2) (L_4 - 2 L_6)
                    + m_K^2 (L_5 - 2 L_8) \big] \: A_{0;2}(m_K^2)
\\ \nonumber && \hphantom{\frac{1}{360 F^4} \{}
+80 \: \big[ (4 m_K^2 - m_\pi^2)^2 L_5 - 48 (m_K^2 - m_\pi^2)^2 L_7
\\ \nonumber && \hphantom{\frac{1}{360 F^4} \{} \hspace{2.75em}
                     + 3 (4 m_K^2 - m_\pi^2) (2 m_K^2 + m_\pi^2) (L_4 - 2 L_6)
\\ \nonumber && \hphantom{\frac{1}{360 F^4} \{} \hspace{2.75em}
                     - 6 (8 m_K^4 - 8 m_K^2 m_\pi^2 + 3 m_\pi^4) L_8
                     \big] \: A_{0;2}(m_\eta^2)
\\ \nonumber && \hphantom{\frac{1}{360 F^4} \{}
+720 \: \big[ 6 m_\pi^2 (L_3 - 2 L_4 + 4 L_1)
      - (2 m_\pi^2 + m_K^2) L_5 \big] \: A_0(m_\pi^2)
\\ \nonumber && \hphantom{\frac{1}{360 F^4} \{}
+1440 m_K^2 \: \big[ 16 L_1 + 4 L_2 + 6 L_3 - 8 L_4 - 3 L_5 \big] \: A_0(m_K^2)
\\ \nonumber && \hphantom{\frac{1}{360 F^4} \{}
+80 \: \big[ 2 (4 m_K^2 - m_\pi^2) (5 L_3 - 6 L_4 + 12 L_1)
     + 3 (2 m_\pi^2 - 7 m_K^2) L_5 \big] \: A_0(m_\eta^2)
\\ \nonumber && \hphantom{\frac{1}{360 F^4} \{}
+45 m_\pi^2 \: A_0(m_\pi^2) \: A_{0;2}(m_\pi^2)
-45 m_\pi^2 \: A_0(m_\pi^2) \: A_{0;2}(m_\eta^2)
\\ \nonumber && \hphantom{\frac{1}{360 F^4} \{}
+120 m_K^2 \: A_0(m_K^2) \: A_{0;2}(m_\eta^2)
-15 m_\pi^2 \: A_0(m_\eta^2) \: A_{0;2}(m_\pi^2)
\\ \nonumber && \hphantom{\frac{1}{360 F^4} \{}
+60 m_K^2 \: A_0(m_\eta^2) \: A_{0;2}(m_K^2)
+5 (7 m_\pi^2-16 m_K^2) \: A_0(m_\eta^2) \: A_{0;2}(m_\eta^2)
\\ \nonumber && \hphantom{\frac{1}{360 F^4} \{}
+45 \: A_0(m_\pi^2)^2
-18 \: A_0(m_K^2)^2
-27 \: A_0(m_\eta^2)^2
\\ \nonumber && \hphantom{\frac{1}{360 F^4} \{}
-54 \: A_0(m_\pi^2) \: A_0(m_\eta^2)
-27 \: A_0(m_\pi^2) \: A_0(m_K^2)
+81 \: A_0(m_K^2) \: A_0(m_\eta^2)
\}
\\ \nonumber &&
+\; ( \delta Z_1^{\text{$K$}} )^2 \:.
\end{eqnarray}

The irreducible diagram yields
\begin{eqnarray}
\delta Z^{\text{$K,\textit{irred}$}}_2 &=&
2.8602989531 
+ \frac{142 m_K^4 + 5 m_K^2 m_\pi^2 + 42 m_\pi^4}
     {72(4\pi)^4 F^4 \, \varepsilon^2}
\\&&
+\; \frac{1}{288(4\pi)^4 F^4 \, \varepsilon} \Big\{
1769 m_K^4 + 186 m_K^2 m_\pi^2 + 624 m_\pi^4
\nonumber\\&&\hphantom{+\; \frac{1}{288(4\pi)^4 F^4 \, \varepsilon} \Big\{}
- 4 ( 56 m_K^4 - 26 m_K^2 m_\pi^2 + 3 m_\pi^4) \: \mathrm{Lg}(m_\eta^2)
\nonumber\\&&\hphantom{+\; \frac{1}{288(4\pi)^4 F^4 \, \varepsilon} \Big\{}
- 24 m_K^2 ( 38 m_K^2 + 3 m_\pi^2) \: \mathrm{Lg}(m_K^2)
\nonumber\\&&\hphantom{+\; \frac{1}{288(4\pi)^4 F^4 \, \varepsilon} \Big\{}
- 36  m_\pi^2 ( 2 m_K^2 + 9 m_\pi^2) \: \mathrm{Lg}(m_\pi^2)
\Big\} \:.
\nonumber
\end{eqnarray}

Finally, the contribution arising from the $\mathcal{L}^{(6)}$-constants of
Eq.~(\ref{3.9c}) are
\begin{eqnarray}
\delta Z_{2}^{\pi }[\mathcal{L}^{(6)}] &=& \frac{1}{F^{4}}\{
\hphantom{+\;}
4m_{\pi}^{4} \: \beta_{17}+4m_{\pi}^{2}(2m_{K}^{2}+m_{\pi}^{2}) \: \beta_{18}
\label{5.13} \\
&&\hphantom{\frac{1}{F^{4}}\{}
+\;4(4m_{K}^{4}-4m_{K}^{2}m_{\pi}^{2}+3m_{\pi}^{4}) \: \beta_{19}
\notag\\
&&\hphantom{\frac{1}{F^{4}}\{}
+\;4(4m_{K}^{4}+4m_{K}^{2}m_{\pi}^{2}+m_{\pi}^{4})
\: \beta_{21}\}  \notag
\\
\delta Z_{2}^{K}[\mathcal{L}^{(6)}] &=& \frac{1}{F^{4}} \{
4(2m_{K}^{4}-2m_{K}^{2}m_{\pi}^{2}+m_{\pi}^{4})\: \beta_{17}
\label{5.14b}
\\
&&\hphantom{\frac{1}{F^{4}}\{}
+4m_{K}^{2}(2m_{K}^{2}+m_{\pi}^{2})\: \beta_{18}
\notag\\
&&\hphantom{\frac{1}{F^{4}}\{}
+4(4m_{K}^{4}-4m_{K}^{2}m_{\pi}^{2}+3m_{\pi}^{4}) \: \beta_{19}
\notag\\
&&\hphantom{\frac{1}{F^{4}}\{}
+4(4m_{K}^{4}+4m_{K}^{2}m_{\pi}^{2}+m_{\pi}^{4})\: \beta_{21}
\} \:. \notag
\end{eqnarray}
If the unrenormalized contributions of the order $p^{2}$, $p^{4}$
and $p^{6}$-diagrams of fig.~2 are denoted as
$\Delta_0 f$, $\Delta_1 f$, and $\Delta_2 f$,
then, with the mass- and wave function renormalizations given above, the
renormalized form factors read
\begin{eqnarray}
f &=&\sqrt{Z^{K}}\;(\Delta_0 f + \Delta_1 f
+ \Delta_2 f )\;\sqrt{Z^{\pi }}%
\;  \label{5.30} \\
&=&\Delta_0 f + \big\{ \Delta_1 f + \Delta_0 f
\big( \frac{1}{2}\delta Z_{1}^{K}+ \frac{1}{2}\delta Z_{1}^{\pi } \big)
\big\}_{_{m^{2}=m_{ph}^{2}-\delta m^2,\;F=F_{\pi }-\delta F}}  \notag \\
&&+\;\Delta_2 f + \Delta_1 f \:
\big\{ \frac{1}{2}\delta Z_{1}^{K}+\frac{1}{2}\delta Z_{1}^{\pi} \big\}
\notag \\
&&+\;\Delta_0 f \:
\big\{ \frac{1}{2}\delta Z_{2}^{K}+\frac{1}{2}\delta Z_{2}^{\pi }
+ \frac{1}{4}\delta Z_{1}^{K}\delta Z_{1}^{\pi }-\frac{1}{8}(\delta
Z_{1}^{K})^{2}-\frac{1}{8}(\delta Z_{1}^{\pi })^{2} \big\}
\notag \\
&& +\; \mathcal{O}(p^{8}) \notag
\end{eqnarray}
where $f$ stands for $f_{\pm}$.


\begin{figure}
  \centerline{
    \includegraphics*[scale=0.8]{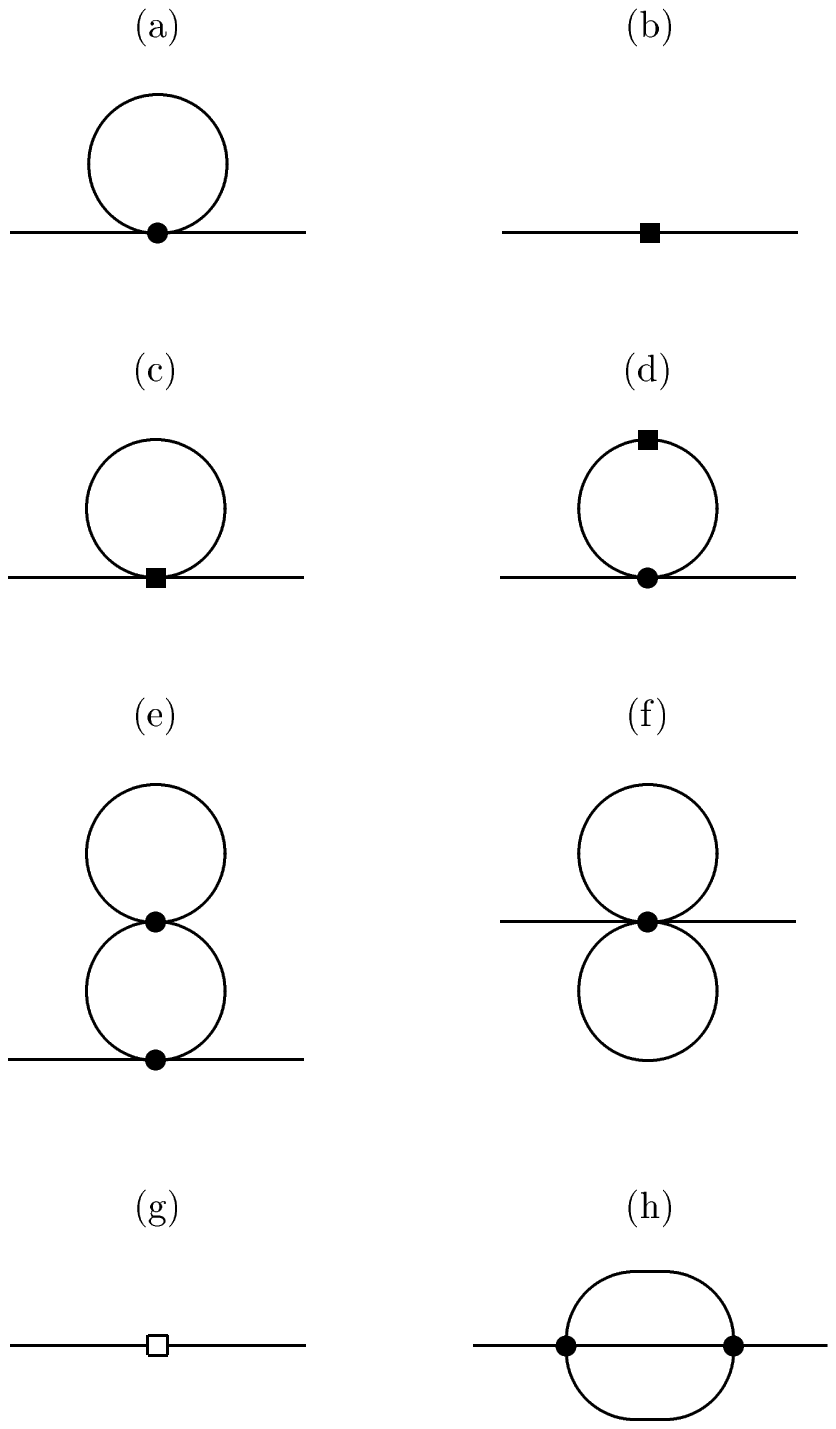}
  }
  Figure 1:
  \footnotesize
  Self energy diagrams up to order $p^{6}$.
   ${\cal L}^{(2)}$-vertices are denoted by filled circles
     (\parbox{1.2ex}{ \unitlength1ex\begin{picture}(0,0)
                   \put(0.5,0){\circle*{1}}\end{picture} }),
  ${\cal L}^{(4)}$-vertices by filled squares
    (\parbox{1ex}{ \unitlength1ex\begin{picture}(0,0)
                   \put(0.5,0){\makebox(0,0){\rule[0ex]{1ex}{1ex}}}
                   \end{picture} }),
  and an ${\cal L}^{(6)}$-vertex by an open square
    (\parbox{1.2ex}{ \unitlength1ex\begin{picture}(0,0)
                     \put(0.5,0){\begin{picture}(0,0)\unitlength0.5ex
                       \put(-1,1){\line(1,0){2}}\put(-1,1){\line(0,-1){2}}
                       \put(1,1){\line(0,-1){2}}\put(-1,-1){\line(1,0){2}}
                     \end{picture}}\end{picture} }).
\end{figure}


\subsection{\protect\bigskip $\mathcal{L}^{(6)}$-contributions to the form
factors}

In every order of chiral perturbation theory there appear new operators with
\emph{a priori} unknown coefficients. The ten constants of $\mathcal{L}%
^{(4)}$ are by now all fixed by experiment, but little is known about the
143 constants of $\mathcal{L}^{(6)}$. Out of the latter, only 11
enter in semileptonic $K$-decay. There are two sources which lead to $%
\mathcal{L}^{(6)}$-contributions to the form factors: one is the $\mathcal{O}%
(p^{6})$ tree graph of fig.~2, and the other one is the
$\mathcal{O}(p^{2})$ tree graph of fig.~2 with the $\mathcal{O}(p^{6})$ wave
function renormalization.
Since to $\mathcal{O}(p^{2})$
\begin{eqnarray}
\;\Delta_0 f_{+}  &=& 1  \label{5.31} \\
\;\Delta_0 f_{-}  &=& 0 \:,  \notag
\end{eqnarray}
the total contribution involving $\mathcal{L}^{(6)}$-constants (ordered in
powers of $q^{2}$) to the form factors $f_{+}$ and $f_{-}$ becomes
\begin{eqnarray}
\Delta f_{+}[\mathcal{L}^{(6)}] &=&\frac{2q^{4}}{F^{4}}\;\{\beta_{22}+\beta
_{23}\}-\frac{2q^{2}}{F^{4}}\{\beta_{22}(m_{K}^{2}+m_{\pi}^{2})-2\beta
_{24}m_{K}^{2}  \label{5.19a} \\
&&-\beta_{25}m_{\pi}^{2}-\beta_{26}(2m_{K}^{2}+m_{\pi}^{2})+\beta
_{27}(m_{K}^{2}+m_{\pi}^{2})\}  \notag \\
&&+\frac{4}{F^{4}}\beta_{14}(m_{K}^{2}-m_{\pi}^{2})^{2}  \notag
\\
\Delta f_{-}[\mathcal{L}^{(6)}] &=&-\frac{2q^{2}}{F^{4}}(m_{K}^{2}-m_{\pi
}^{2})\{\beta_{8}+\beta_{22}+\beta_{23}\}  \label{5.19b} \\
&&+\;\frac{m_{K}^{2}-m_{\pi}^{2}}{F^{4}}\{2\beta_{8}(m_{K}^{2}+m_{\pi
}^{2}) - 2\beta_{16}(2m_{K}^{2} + m_{\pi}^{2})
-4\beta_{17}m_{K}^{2}  \notag
\\
&& - 2\beta_{18}(2m_{K}^{2} + m_{\pi}^{2})+2\beta_{22}(m_{K}^{2}+m_{\pi
}^{2})-4\beta_{24}m_{K}^{2}  \notag \\
&&-2\beta_{25}m_{\pi}^{2} - 2\beta_{26}(2m_{K}^{2} + m_{\pi}^{2})+2\beta
_{27}(m_{K}^{2}+m_{\pi}^{2})\}  \notag
\end{eqnarray}

The $q^{4}$-term of $\Delta f_{+}[\mathcal{L}^{(6)}]$,
i.e. $2q^4(\beta_{22}+\beta_{23})/F^4$, is the same as that for the electromagnetic
form factor of the charged pion (and kaon). One can therefore use data on the second
derivative of the pion electromagnetic form factor at the origin, together
with the $\mathcal{O}(p^{6})$ loop calculation, to determine the combination
$\beta_{22}+\beta_{23}$. Details are given in appendix \ref{pionFFdiscussion}.
We find for the $\varepsilon^0$-part
\begin{equation}
(\beta_{22}+\beta_{23})^{(0),\text{\it GL}}=0.61\times 10^{-4} \label{5.19c}
\end{equation}
where we have used the \emph{GL}-scheme (\ref{betaGL}) and chosen
the mass scale $\mu=770$~MeV.
To these counter term contributions, we have to add those of the loop
diagrams involving $\mathcal{L}^{(2)}$- and $\mathcal{L}^{(4)}$-vertices.


\begin{figure}
  \centerline{
    \includegraphics*[width=5.7294in]{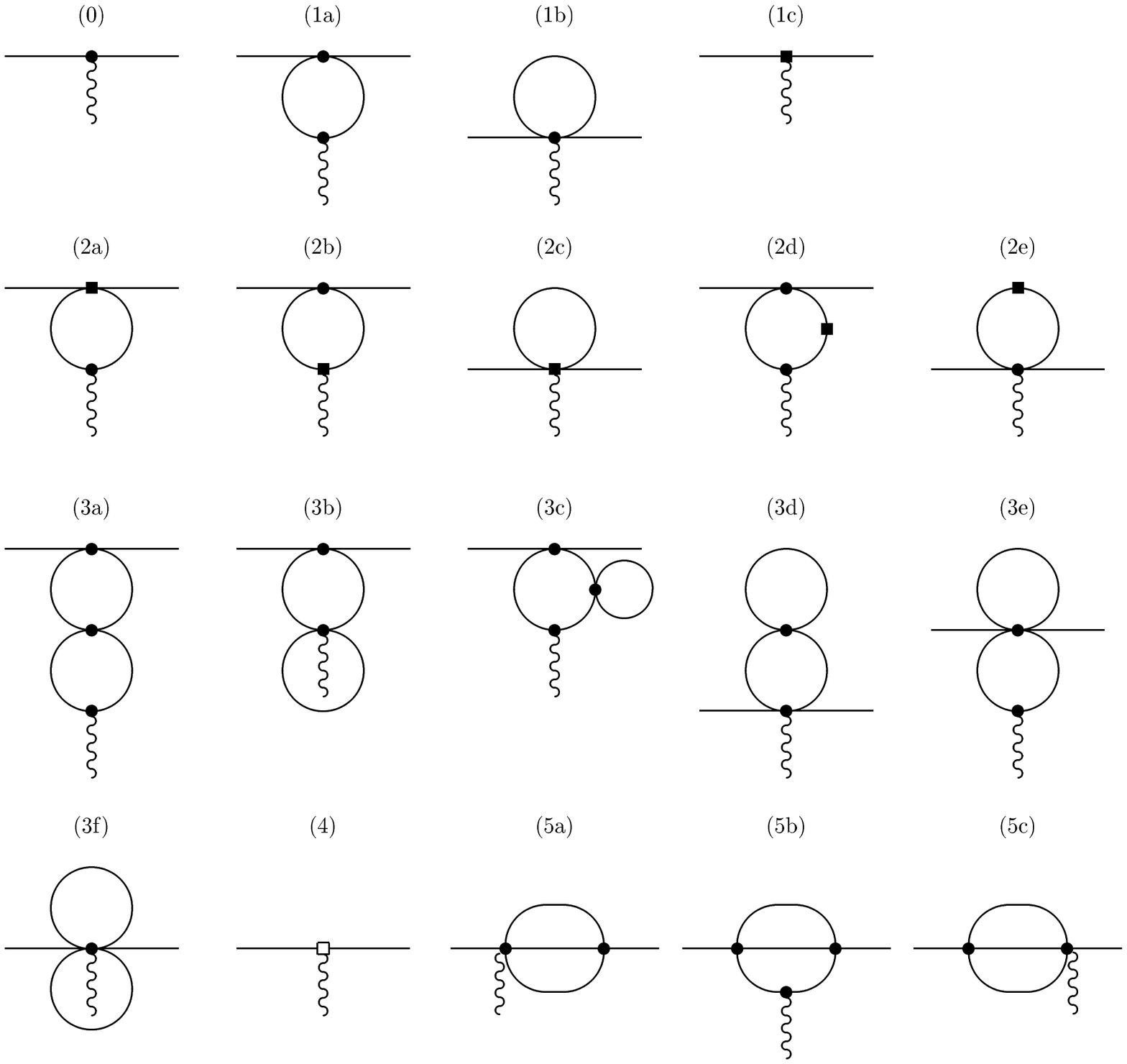}
  }
  Figure 2:
  \footnotesize
  The diagrams for the $K_{l3}$ form factor up to order $p^{6}$.
  ${\cal L}^{(2)}$-vertices are denoted by filled circles
     (\parbox{1.2ex}{ \unitlength1ex\begin{picture}(0,0)
                   \put(0.5,0){\circle*{1}}\end{picture} }),
  ${\cal L}^{(4)}$-vertices by filled squares
    (\parbox{1ex}{ \unitlength1ex\begin{picture}(0,0)
                   \put(0.5,0){\makebox(0,0){\rule[0ex]{1ex}{1ex}}}
                   \end{picture} }),
  and an ${\cal L}^{(6)}$-vertex by an open square
    (\parbox{1.2ex}{ \unitlength1ex\begin{picture}(0,0)
                     \put(0.5,0){\begin{picture}(0,0)\unitlength0.5ex
                       \put(-1,1){\line(1,0){2}}\put(-1,1){\line(0,-1){2}}
                       \put(1,1){\line(0,-1){2}}\put(-1,-1){\line(1,0){2}}
                     \end{picture}}\end{picture} }).
  Diagrams (0) to (3f) are referred to as ''reducible'',
  diagrams (5a) to (5c) as ''irreducible''.
\end{figure}


\subsection{Divergent two-loop contributions to $f_{\pm}$}

We start with an analysis of the divergent two-loop contributions of fig.~2
to the form factors. The $\mathcal{O}(p^{6})$ pole terms in $\varepsilon $ have
to cancel in the sum of all loops and the tree graphs with an $\mathcal{L}^{(6)}$%
-vertex:
\begin{equation}\label{5.21}
\Big( \Delta_{p^6}^{\text{\it loop}} f_{\pm}
+\Delta f_{\pm}[\mathcal{L}^{(6)}] \Big)_{\text{\it div}} \;=\; 0 \:.
\end{equation}
Since the $\mathcal{L}^{(6)}$-tree graph part $\Delta f_{\pm}[\mathcal{L}^{(6)}]$
is a polynomial in masses and momenta, it follows that the loop part
$\Delta_{p^6}^{\text{\it loop}} f_{\pm}$ must also be
polynomial in masses and momenta, i.e. it cannot
contain any logarithms thereof. This condition offers a good check of the
calculation. In fact, we find that in the sum of all
$\mathcal{O}(p^6)$ loop diagrams any non-polynomial terms in masses and
momenta cancel:
\begin{eqnarray}
\Delta_{p^6}^{\text{\it loop}} f_{+} &=&\frac{1}{96(4\pi F)^{4}\varepsilon ^{2}}%
\{21m_{K}^{4}-42m_{K}^{2}m_{\pi}^{2}-34m_{K}^{2}q^{2}+21m_{\pi
}^{4}-28m_{\pi}^{2}q^{2}-3q^{4}\}  \label{5.22} \\
&&+\;\frac{1}{576(4\pi )^{2}F^{4} \varepsilon }%
\{768\, L_{1}^{(0)}(-m_{K}^{4}+2m_{K}^{2}m_{\pi}^{2}+3m_{K}^{2}q^{2}-m_{\pi
}^{4}+3m_{\pi}^{2}q^{2}-q^{4})  \notag \\
&&\qquad +\;384\, L_{2}^{(0)}(-m_{K}^{4}+2m_{K}^{2}m_{\pi}^{2}-3m_{K}^{2}q^{2}-m_{\pi
}^{4}-3m_{\pi}^{2}q^{2}+q^{4})  \notag \\
&&\qquad +\;64\, L_{3}^{(0)}(-m_{K}^{4}+2m_{K}^{2}m_{\pi}^{2}+45m_{K}^{2}q^{2}-m_{\pi
}^{4}+9m_{\pi}^{2}q^{2}-9q^{4})  \notag \\
&&\qquad +\;768\, L_{4}^{(0)}(3m_{K}^{4}-6m_{K}^{2}m_{\pi}^{2}+m_{K}^{2}q^{2}+3m_{\pi
}^{4}+m_{\pi}^{2}q^{2})  \notag \\
&&\qquad +\;576\, L_{5}^{(0)}(-3m_{K}^{4}+6m_{K}^{2}m_{\pi
}^{2}+m_{K}^{2}q^{2}-3m_{\pi}^{4}+m_{\pi}^{2}q^{2})  \notag \\
&&\qquad +\;288\, L_{9}^{(0)}q^{2}(7m_{K}^{2}+5m_{\pi}^{2}+q^{2})  \notag \\
&&\qquad -\;\frac{1}{(4\pi )^{2}}[70m_{K}^{4}-140m_{K}^{2}m_{\pi
}^{2}+167m_{K}^{2}q^{2}+70m_{\pi}^{4}+125m_{\pi}^{2}q^{2}-12q^{4}]\}%
\notag
\end{eqnarray}
and%
\begin{eqnarray}
\Delta_{p^6}^{\text{\it loop}} f_{-} &=&\frac{1}{48(4\pi F)^{4}\varepsilon ^{2}}%
\{29m_{K}^{4}+18m_{K}^{2}m_{\pi}^{2}-27m_{K}^{2}q^{2}-47m_{\pi
}^{4}+27m_{\pi}^{2}q^{2}\}  \label{5.23} \\
&&+\;\frac{1}{576(4\pi )^{2}F^{4} \varepsilon }%
\{768\, L_{1}^{(0)}(-7m_{K}^{4}+3m_{K}^{2}q^{2}+7m_{\pi}^{4}-3m_{\pi
}^{2}q^{2})  \notag \\
&&\qquad +\;384\, L_{2}^{(0)}(-19m_{K}^{4}+7m_{K}^{2}q^{2}+19m_{\pi
}^{4}-7m_{\pi}^{2}q^{2})  \notag \\
&&\qquad +\;64\, L_{3}^{(0)}(-145m_{K}^{4}+48m_{K}^{2}m_{\pi
}^{2}+41m_{K}^{2}q^{2}+97m_{\pi}^{4}-41m_{\pi}^{2}q^{2})  \notag \\
&&\qquad +\;384\, L_{4}^{(0)}(-26m_{K}^{4}+9m_{K}^{2}m_{\pi}^{2}+17m_{\pi
}^{4})  \notag \\
&&\qquad +\;64\, L_{5}^{(0)}(-m_{K}^{4}+2m_{K}^{2}m_{\pi
}^{2}+27m_{K}^{2}q^{2}-m_{\pi}^{4}-27m_{\pi}^{2}q^{2})  \notag \\
&&\qquad +\;6912\, L_{6}^{(0)}(2m_{K}^{4}-m_{K}^{2}m_{\pi}^{2}-m_{\pi}^{4})
\notag \\
&&\qquad +\;9216\, L_{7}^{(0)}(m_{K}^{4}-2m_{K}^{2}m_{\pi}^{2}+m_{\pi}^{4})
\notag \\
&&\qquad +\;2304\, L_{8}^{(0)}(5m_{K}^{4}-4m_{K}^{2}m_{\pi}^{2}-m_{\pi}^{4})
\notag \\
&&\qquad +\;288\, L_{9}^{(0)}(-7m_{K}^{4}+2m_{K}^{2}m_{\pi
}^{2}-m_{K}^{2}q^{2}+5m_{\pi}^{4}+m_{\pi}^{2}q^{2})  \notag \\
&&\qquad +\;\frac{1}{(4\pi )^{2}}[11m_{K}^{4}-204m_{K}^{2}m_{\pi
}^{2}-51m_{K}^{2}q^{2}+193m_{\pi}^{4}+51m_{\pi}^{2}q^{2}]\} \:.  \notag
\end{eqnarray}
This is \emph{not} the case for the group of reducible resp. the group of
irreducible diagrams alone, only in their sum.
$L_i^{(0)}$ are the $\varepsilon^0$-coefficients of the $\mathcal{L}^{(4)}$ constants,
cf. eq.~(\ref{5.41b}).

The divergent parts $\Delta f_{\pm}[\mathcal{L}^{(6)}]$, given explicitly
in terms of the $\mathcal{L}^{(6)}$-constants $\beta_{j}$
in eqs.~(\ref{5.19a}) and (\ref{5.19b}), have to be the negative of these
expressions so that the whole $\mathcal{O}(p^6)$ prediction is finite.
Since the $\mathcal{L}^{(6)}$-constants $\beta_j$ do not know anything
about the masses $m_\pi^2$, $m_K^2$, they must appear in (\ref{5.19a})
and (\ref{5.19b}) in such a way that the mass dependence of the divergent parts
$\Delta f_{\pm}[\mathcal{L}^{(6)}]$ is produced from the explicit
masses in (\ref{5.19a}), (\ref{5.19b}) alone. In other words:
the $\mathcal{L}^{(6)}$-constants themselves cannot contribute any masses.
In appendix \ref{divPartsOfBeta}
we list the resulting divergent parts for the relevant $\mathcal{L}^{(6)}$-constants
$\beta_j$. The fact that they are independent of the masses, is a
consistency check between our calculation and the Lagrangian $\mathcal{L}^{(6)}$
from \cite{FearingScherer}.


\subsection{Reducible loop diagrams}

The reducible diagrams of fig.~2 can be expressed in terms of
one-loop integrals. In case of two loops they are of the form
\begin{equation}
\mu ^{8-2D}\int \frac{d^{D}k}{(2\pi )^{D}}\;\frac{d^{D}l}{(2\pi )^{D}%
}\;\frac{V(k^{2},kp_{1},kp_{2},kl,l^{2},lp_{1},lp_{2})}{P}  \label{6.1}
\end{equation}
and in case of one loop
\begin{equation}
\mu ^{4-D}\int \frac{d^{D}k}{(2\pi )^{D}}\;\frac{V(k^{2},kp_{1},kp_{2})}{%
P} \:. \label{6.2}
\end{equation}
Here $V$ is a polynomial of its arguments and $P$
represents a product of propagator factors that depends on the topology of
the diagram.
A first simplification of these integrals is achieved by replacing
certain factors in the numerator according to
\begin{eqnarray}
k^{2} &=&P(k,m^{2})+m^{2}  \label{6.3} \\
kq &=&\frac{1}{2}[P(k+q,m^{2})+m^{2}-k^{2}-q^{2}]  \notag \\
l^{2} &=&P(l,m^{2})+m^{2}  \notag \\
lq &=&\frac{1}{2}[P(l+q,m^{2})+m^{2}-l^{2}-q^{2}]\;,  \notag
\end{eqnarray}
where $P(k,m^{2})=k^{2}-m^{2}+i0$. The
remaining integrals can be expressed through the 1-loop one-point function $%
A_{0}(m^{2})$ of eq. (\ref{5.7a1}), the 1-loop two-point function
\begin{equation}
B_0(q^2,m_1^2,m_2^2) \;=\;
\mu^{4-D}\int \frac{d^{D}k}{i(2\pi )^{D}}\;\frac{1}{%
[(k+q)^{2}-m_{1}^{2}][k^{2}-m_{2}^{2}]} \:,
\label{6.4}
\end{equation}
and tensors and mass derivatives thereof.

In a two-loop calculation these functions have to be considered up to order
$\varepsilon^1$ (where $D=4-2\varepsilon$).
The results are given in appendix \ref{OneLoopIntegrals}.
The reducible contributions to $f_+$ are given explicitly
for each diagram in appendix \ref{redContribToFPlus}.


\section{Irreducible two-loop diagrams}

In the irreducible diagrams 5a, 5b, 5c of fig.~2 the two
loop integrations are not independent of each other as they were in the reducible
graphs. That is why genuine 2-loop functions enter the stage which cannot be
expressed by 1-loop integrals only.

Inserting the Feynman rules yields integrands with a similar structure as in
(\ref{6.1})
\begin{equation}
\frac{V(k^{2},kp_{1},kp_{2},kl,l^{2},lp_{1},lp_{2})}{%
P(k+p_{1},m_{0}^{2})P(k+p_{2},m_{1}^{2})P(k+l,m_{2}^{2})P(l,m_{3}^{2})},
\label{6.9}
\end{equation}
where $V$ is a polynomial of degree equal to the number of vertices. After
canceling factors via
\begin{eqnarray}
kp_{1} &=&\frac{1}{2}[P(k+p_{1},m_{1}^{2})+m_{0}^{2}-k^{2}-p_{1}^{2}]
\label{6.10} \\
kp_{2} &=&\frac{1}{2}[P(k+p_{2},m_{1}^{2})+m_{1}^{2}-k^{2}-p_{2}^{2}]  \notag
\\
kl &=&\frac{1}{2}[P(k+l,m_{2}^{2})+m_{2}^{2}-k^{2}-l^{2}]  \notag \\
l^{2} &=&P(l,m_{3}^{2})+m_{3}^{2}  \notag
\end{eqnarray}
we are left with reducible integrals which can be calculated analytically,
and with some genuine 2-loop integrals of the \emph{sunset}-topology, i.e.
the 3-point functions
\begin{eqnarray}
T_{\alpha_{1},\alpha_{2},\beta
}(q^{2};p_{1}^{2},p_{2}^{2};m_{0}^{2},m_{1}^{2},m_{2}^{2},m_{3}^{2}) &=&
\notag \\
&&\hspace{-13em}\mu ^{8-2D}\int \frac{d^{D}k}{i(2\pi )^{D}}\;\frac{%
d^{D}l}{i(2\pi )^{D}}\;\frac{(lp_{1})^{\alpha_{1}}\;(lp_{2})^{\alpha
_{2}}\;(k^{2})^{\beta }}{P(k+p_{1},m_{0}^{2})\;P(k+p_{2},m_{1}^{2})%
\;P(k+l,m_{2}^{2})\;P(l,m_{3}^{2})}\quad \hspace{3em}
\end{eqnarray}
and the 2-point functions
\begin{equation}
S_{\alpha ,\beta }(p^{2};m_{1}^{2},m_{2}^{2},m_{3}^{2})=\mu ^{8-2D}\int
\frac{d^{D}k}{i(2\pi )^{D}}\;\frac{d^{D}l}{i(2\pi )^{D}}\;\frac{%
(lp)^{\alpha }\;(k^{2})^{\beta }}{P(k+p,m_{1}^{2})\;P(k+l,m_{2}^{2})%
\;P(l,m_{3}^{2})}.\hspace{3em}  \label{6.12}
\end{equation}

In diagram (5b) of fig.~2, nine different mass flows of intermediate
mesons must be regarded:
\begin{eqnarray}
\Diag{5b}{}{}{\pm} &=& \hphantom{+\;}
                           \Diag{5b}{K\pi\pi\pi}{}{\pm}
                       +\; \Diag{5b}{K\pi KK}{}{\pm}
                       +\; \Diag{5b}{K\pi\eta\eta}{}{\pm}
         \\ &&         +\; \Diag{5b}{\pi K\pi K}{}{\pm}
                       +\; \Diag{5b}{\pi KK\eta}{}{\pm}
                       +\; \Diag{5b}{\eta K\pi K}{}{\pm}
\nonumber\\ &&         +\; \Diag{5b}{\eta KK\eta}{}{\pm}
                       +\; \Diag{5b}{K\eta KK}{}{\pm}
                       +\; \Diag{5b}{K\eta\pi\eta}{}{\pm}
\end{eqnarray}
where $\Diag{5b}{rstu}{}{\pm}$ means that mesons $r$ and $s$ couple
to the $W$-boson and the other two lines are mesons of type $t$ and $u$.
Each mass flow is handled separately, and its contribution to the $K_{l3}$ form factor is
expressed in terms of the basic 1- and 2-loop functions $A$, $B$, $S_{\alpha
,\beta }$, $T_{\alpha_{1},\alpha_{2},\beta }$, where for the latter at
most the tensor indices
\begin{quote}
$S_{2,0}$, $S_{1,1}$, $S_{1,0}$, $S_{0,0}$, $S_{0,1}$, $S_{0,2}$, $T_{0,0,0}$%
, $T_{0,0,1}$, $T_{0,0,2}$, $T_{0,0,3}$, $T_{1,0,0}$, $T_{1,0,1}$, $%
T_{1,0,2} $, $T_{1,1,0}$, $T_{1,1,1}$
\end{quote}
are needed. Except for special kinematic situations the genuine 2-loop
integrals $S_{\alpha ,\beta }$ and $T_{\alpha_{1},\alpha_{2},\beta }$
cannot be calculated analytically. In \cite{PostSchilcher} we describe the
method how we calculated them by splitting them up into one part which contains
the divergence and can be evaluated analytically, and a second part which is
finite and can be done numerically.


\section{Finite contributions of the loops}

After presenting the results of the pole-terms of the loop-diagrams we now come
to their finite parts which contain the actual physical information. We will
present the results, which can only be given in numerical form, graphically
and as interpolation polynomials. We use the $GL$-scheme
discussed above at a scale $\mu =m_{\rho }=770$~MeV, the masses given in
(\ref{massenwerte}), and the following values for the finite parts
$L_i^{\text{\it ren}}$ of the $\mathcal{L}^{(4)}$-constants:
\addtolength{\tabcolsep}{-1mm}
\begin{equation}\label{Tabelle_der_Lren}
\parbox{14.9cm}{
  \begin{tabular}{r|cccccccccc}
  $i$        & 1 & 2  & 3     & 4    & 5  & 6    & 7    & 8 & 9
  \\ \hline \rule{0mm}{2.5ex}
  $10^4\,L_i^{\mbox{\scriptsize\em ren}}$
             & $7\pm 5$ & $12\pm 4$ & $- 35\pm 13$ & $- 3\pm 5$ & $14\pm 5$ & $- 2\pm 3$ & $- 4\pm 2$ & $9\pm 3$ & $69\pm 7$
\end{tabular}}
\end{equation}
\addtolength{\tabcolsep}{1mm}
The $\varepsilon^1$-coefficients of $L_i$ are not new degrees of freedom,
but always appear in combination with certain $\mathcal{L}^{(6)}$-parameters.
Therefore, we define
\begin{equation}
L_i^{(1), GL}(\mu = m_\rho) \;=\; 0 \:,
\end{equation}
cf. (\ref{5.44}). As our kinematical
range we choose momentum transfers $-m_K^{2}\leq q^{2}\leq
(m_{K}-m_{\pi})^{2}$, which is certainly within the range of applicability of
chiral perturbation theory. The results of the $\mathcal{O}(p^{4})$- and the
$\mathcal{O}(p^{6})$-loop contributions (reducible and irreducible) for the
form factors $f_{+}(q^2)$ and $f_-(q^2)$ are plotted as a function of $x=q^{2}/m_{K}^{2}$ \
in figs.~3 and~4. We observe that for $f_+$ the reducible and the irreducible
$\mathcal{O}(p^6)$-contributions cancel almost completely. For $f_-$ the irreducible loop
corrections are very small. For both $f_+$ and $f_-$ the $\mathcal{O}(p^6)$
loop corrections are essentially linear functions of $q^2$, i.e. they 
create only small nonlinear contributions to the form factors.


\begin{figure}
  \centerline{ 
  \includegraphics*[width=5.7553in]{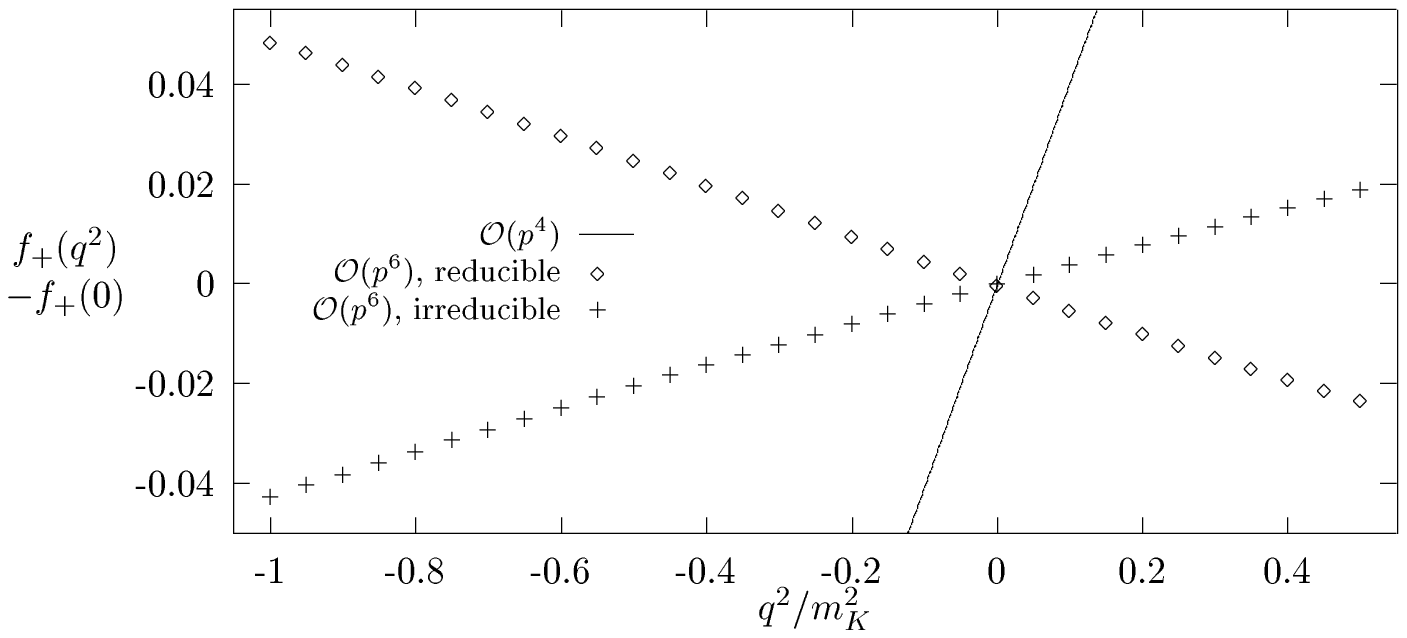} } 
  Figure 3: 
  \footnotesize Loop contributions to the form factor $f_{+}(q^{2})$.
  The solid line is the leading order $\mathcal{O}(p^4)$ result, the
  dotted curves denote the corrections at order $p^6$ due to the
  reducible resp.\ irreducible loop diagrams in the $GL$-scheme (on the
  one hand diagrams (2a)--(3f) and the $p^6$ contributions of the
  renormalized leading order diagrams (0), (1a)--(1c), on the other hand
  diagrams (5a)--(5c), cf. fig.~2).  The contribution of diagram (4) is
  missing: it would add an arbitrary parabola through the origin.
\end{figure}


\begin{figure}
  \centerline{
    \includegraphics*[width=5.7553in]{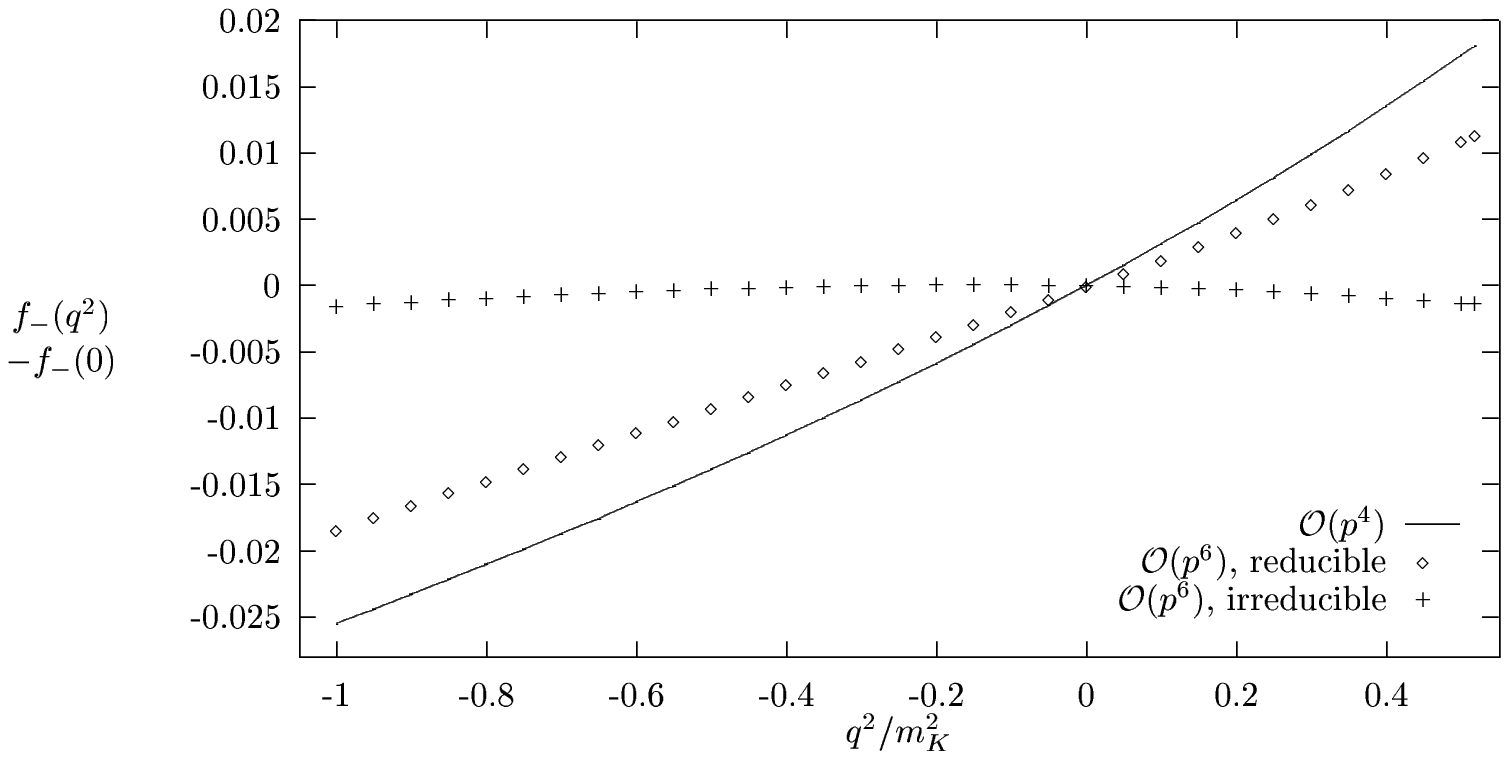}
  }
  Figure 4:
  \footnotesize
  Loop contributions to the form factor $f_{-}(q^{2})$,
  in analogy to fig.~3. The contribution from the $\mathcal{L}^{(6)}$-tree graph
  (4) is missing. It would add an arbitrary straight line through the origin.
\end{figure}

The interpolation polynomials for the loop corrections at order $p^6$ of
the form factors $f_{\pm \text{ }}$ (i.e. the $\mathcal{O}(p^4,p^6)$
parts of the full renormalized form factors given in eq.~(\ref{5.30}),
but without the tree graphs from $\mathcal{L}^{(6)}$) read
\begin{align}
\Delta_{p^{6}}^{\mbox{\scriptsize\em loop}} f_{+} &= x(-0.0101+0.0009%
\,x+0.0054\,x^{2}+0.0007\,x^{3})+\Delta_{p^{6}}^{\mbox{\scriptsize\em loop}%
} f_{+}(0) \label{7.1a}
\\
\Delta_{p^{6}}^{\mbox{\scriptsize\em loop}} f_{-}
&= x(0.0185+0.0001\,x+0.0022\,x^{2}+0.0008\,x^{3})+\Delta_{p^{6}}^{%
\mbox{\scriptsize\em loop}} f_{-}(0) \:,\label{7.4}
\end{align}
$x = q^2/m_K^2 \in [0,0.5]$.
For completeness, we quote the results for $f_{\pm}(0)$ which
come from the $\mathcal{O}(p^{4})$ and $\mathcal{O}(p^{6})$ loops:
\begin{eqnarray}
\Delta^{\mbox{\scriptsize\em loop}}_{p^4+p^6}
f_{+}(0)  &=&\Delta_{p^{4}} f_{+}(0) +
\Delta_{p^{6}}^{\mbox{\scriptsize\em red. loop}%
} f_{+}(0) +\Delta_{p^{6}}^{\mbox{\scriptsize\em irred. loop}} f_{+}(0) \qquad
\label{7.5} \\
&=& -0.0229 + 0.00937 + 0.00853 = -0.0050 \notag \\
\Delta ^{\mbox{\scriptsize\em loop}}_{p^4+p^6}  f_{-}(0) &=&
\Delta_{p^{4}} f_{-}(0)  + \Delta_{p^{6}}^{\mbox{\scriptsize\em red. loop}%
} f_{-}(0) + \Delta_{p^{6}}^{\mbox{\scriptsize\em irred. loop}} f_{-}(0) \qquad
\label{7.7} \\
&=&-0.1836 + 0.0832 - 0.0533 = -0.1537 \:. \notag
\end{eqnarray}
Here, $\Delta_{p^{6}}^{\mbox{\scriptsize\em irred. loop}}$ denotes the
contributions of the irreducible diagrams (5a), (5b), (5c) from fig.~2
and the irreducible part of the wave function renormalization (diagram
(h) from fig.~1). $\Delta_{p^{6}}^{\mbox{\scriptsize\em red. loop}}$
stands for the remaining loop contributions which all come from
reducible diagrams.  To these results, the contributions of the
$\mathcal{L}^{(6)}$ constants, which occur at $\mathcal{O}(p^6)$, must
be added. These can either be determined by experiment or by model
calculations.

The $\mathcal{O}({p^6})$ loop contributions still depend on the mass
scale $\mu$.  This dependence follows from their divergent parts which
are given in eq.~(\ref{5.22}). For $f_+(0)$ we find a value of $0.0617$
at the scale $\mu=500$~MeV and a value of $-0.0364$ at the scale
$\mu=1000$~MeV to be compared with the above value $-0.0050$ at
$\mu=770$~MeV. It is seen that the loop contribution to $f_{\pm}(0)$ is
rather sensitive to the choice of $\mu$ due to cancellations of the
reducible and the irreducible contributions. This $\mu$-dependence is of
course canceled by the $\mu$-dependence of the $q^6$ counter terms. Eq.\ 
(\ref{7.5}) and (\ref{7.7}) only serve as an indication of the size of
the effects to be expected.

At the present stage the predictive power of our $\mathcal{O}(p^{6})$
calculation lies only in quantifying a deviation from linear form factor
rise.  From eqs. (\ref{5.19a}) and (\ref{5.19c}) we find a nonlinear
contribution to $f_+$ of
\begin{equation}\label{quadrfPlusContrib}
  \Delta^{\text{\it nonlin.}}_{p^6} f_+(q^2) \;=\; 0.10 \; \frac{q^4}{m_K^4} \:,
\end{equation}
whereas $f_-$ contains negligible nonlinearities in the relevant
kinematic range.  The effect of the quadratic term in $f_+$ is
essentially a lowering of the parameter $\lambda_+$ defined in
(\ref{2.4}).  In our fit we find
\begin{equation}
\lambda_+ \;=\; 0.022
\end{equation}
(cf. fig.~5) as compared to the linear fit $\lambda_+ = 0.0245$ in \cite{apostolakis}.

We conclude this section with a short discussion of the errors of the $%
\mathcal{O}(p^{6})$ correction. Apart from the general ambiguity due to the $%
\mathcal{L}^{(6)}$ counter terms, errors arise from the $\mathcal{L}^{(4)}$
constants which appear in the $\mathcal{O}(p^{6})$ corrections, and from the one $%
\mathcal{L}^{(6)}$ constant, associated with the $q^{4}$ term of $f_{+}(q^{2})$,
which can be extracted from the pion
electromagnetic form factor. Although these errors could be as large as
10\%, they are irrelevant to our result as the total $\mathcal{O}(p^{6})$
effect is small, see fig.~5.


\begin{figure}
  \centerline{
    \includegraphics*[width=5.8237in]{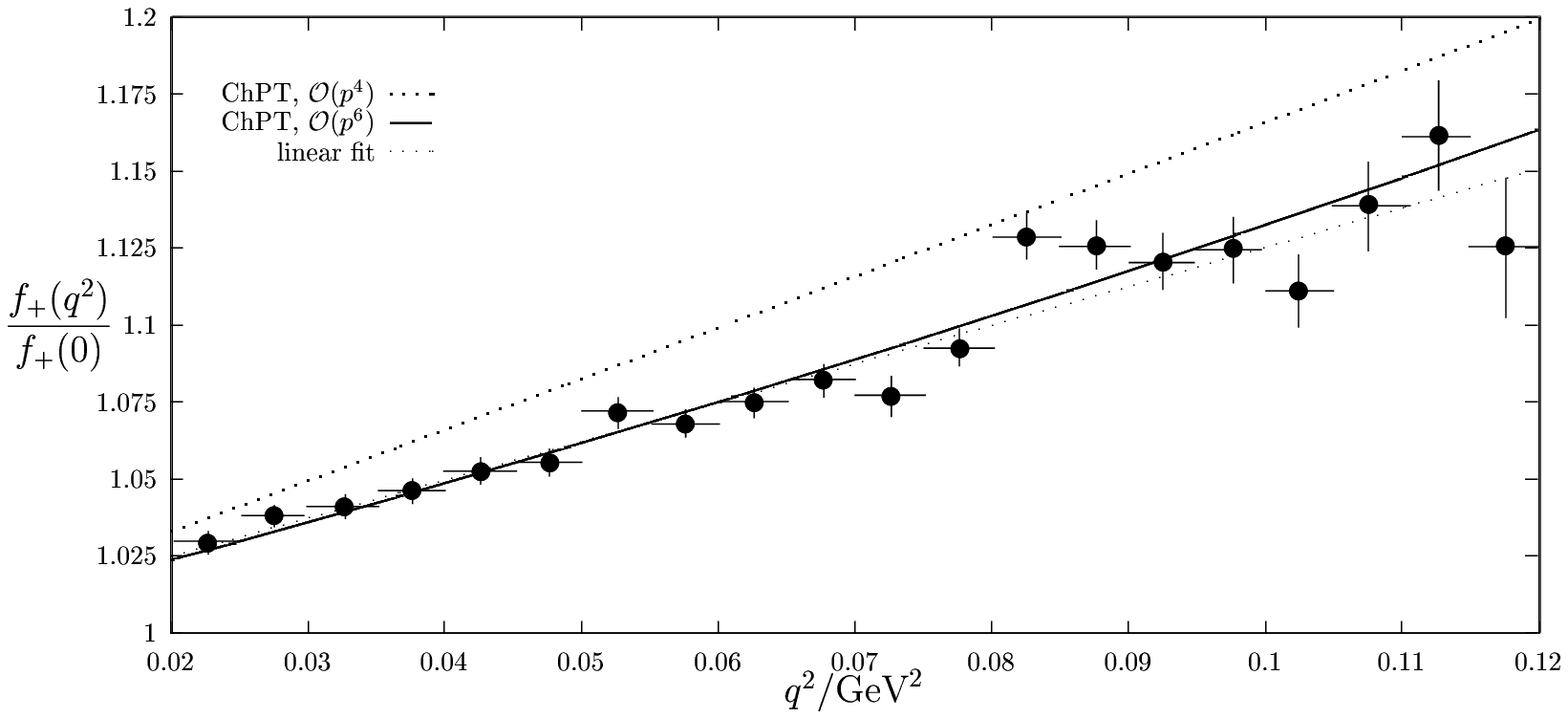}
  }
  Figure 5:
  \footnotesize
  The kaon semileptonic form factor $f_+$ to $\mathcal{O}(p^{4})$ and
  $\mathcal{O}(p^{6})$ chiral perturbation theory versus experiment.
  The data are from the CPLEAR experiment \cite{apostolakis},
  the upper dotted curve is the leading order $\mathcal{O}(p^4)$-result
  of chiral perturbation theory,
  the solid curve is the full result at order $p^6$,
  and the lower dotted curve is the linear fit from \cite{apostolakis}.
  The slope at the origin of the $\mathcal{O}(p^{6})$ result has been fitted
  to the data, its second derivative comes from the electromagnetic
  pion form factor.
\end{figure}


\section{Analysis of results and conclusions}

We have calculated the $\mathcal{O}(p^{6})$ contribution to semileptonic
$K_{l3}$ decay in $SU(3)\times SU(3)$ chiral perturbation theory. This
is an effective field theory, so that there appear new operators with
unknown couplings in each order of perturbation theory. For the $K_{l3}$
form factors $f_{\pm}(q^{2})$ this means that the constant and the
linear term in $q^{2}$ are not determined by the theory. The
$q^{4}$-counter term, however, is the same for the semileptonic form
factor $f_{+}(q^{2})$ and the electromagnetic form factor of the charged
pion. It can therefore be extracted by comparing the
$\mathcal{O}(p^{6})$ chiral perturbation theory with low energy data on
the pion electromagnetic form factor. The details are described in
appendix \ref{pionFFdiscussion}. There is no $q^{4}$-counter term for
the form factor $f_{-}(q^{2})$.

We have found that the $\mathcal{O}(p^6)$ loop corrections are essentially
linear in $q^2$. Thus, the only nonlinearity is the $q^4$-contribution
to $f_+$ coming from the $\mathcal{L}^{(6)}$ tree graph which is related
to the electromagnetic pion form factor.

It is interesting to consider the reducible and the irreducible
$\mathcal{O}(p^{6})$ loop results separately as plotted in figs.~3
and~4. It is clearly seen that for $f_+$ both terms cancel almost
exactly in the physical range of $K_{l3}$ decay. It should be kept in
mind, however, that an arbitrary linear term can always be added to the
$\mathcal{O}(p^{6})$ predictions. For $f_-$ the irreducible
contributions are very small.

The results obtained are interesting from the following points of view:
\begin{itemize}
\item The convergence of chiral perturbation theory in semileptonic $K_{l3}$%
      -decay is established. It turns out that in $K_{l3}$ decay the
      $\mathcal{O}(p^{6})$ 
      corrections are small. This need not always be the case
      \cite{PostSchilcher}.
\item For the form factors, the deviation from a linear rise in $q^{2}$ is
      small, but not negligible (for $f_+$). The result of the nonlinear
      contribution to $f_+$ is effectively a
      lowering of the parameter $\lambda_+$.
\item Our method of calculating the irreducible two-loop diagrams of high
      tensorial rank and involving three different masses can also
      be applied elsewhere.
\item The wave function renormalization constants calculated here can be
      employed in other processes.
\item The divergent parts of the relevant $\mathcal{L}^{(6)}$-parameters
      $\beta_j$ appear in other processes as well and can be used there
      as a check.
\end{itemize}

\bigskip

\bigskip



\begin{appendix}

\section{1-Loop integrals}
\label{OneLoopIntegrals}

In this appendix we reproduce the well known one-loop results. With the
definition
\begin{equation}
\Lg(m^{2}) \;=\; \log (\frac{m^{2}}{4\pi \mu ^{2}})+\gamma
\label{A.1}
\end{equation}
one has
\begin{eqnarray}
A_{0}(m^{2}) &=&\mu ^{4-D}\int \frac{d^{D}k}{i(2\pi )^{D}}\;\frac{1}{%
k^{2}-m^{2}} \label{A.2} \\
&=&-\frac{\mu ^{2}}{4\pi }\;\Gamma (-1+\varepsilon )\;(\frac{m^{2}}{4\pi \mu
^{2}})^{1-\varepsilon }  \notag \\
&=&\frac{m^{2}}{(4\pi )^{2}}\{\frac{1}{\varepsilon }+1-\Lg(m^{2})+\varepsilon
\lbrack 1+\frac{\pi ^{2}}{12}-\Lg(m^{2})+\frac{\Lg(m^{2})^{2}}{2}]+\mathcal{O}%
(\varepsilon ^{2})\}\;,\quad  \notag
\end{eqnarray}
and
\begin{eqnarray}
B_{0}(q^{2};m_{1}^{2},m_{2}^{2}) &=&\mu ^{4-D}\int \frac{d^{D}k}{i(2\pi
)^{D}}\;\frac{1}{[(k+q)^{2}-m_{1}^{2}][k^{2}-m_{2}^{2}]}  \label{A.3} \\
&=&\frac{b^{(-1)}(q^{2};m_{1}^{2},m_{2}^{2})}{\varepsilon }%
+b^{(0)}(q^{2};m_{1}^{2},m_{2}^{2})+\varepsilon
\;\!b^{(1)}(q^{2};m_{1}^{2},m_{2}^{2})+\mathcal{O}(\varepsilon ^{2}),\qquad
\notag
\end{eqnarray}
with%
\begin{eqnarray}
b^{(-1)}(q^{2};m_{1}^{2},m_{2}^{2}) &=&\frac{1}{(4\pi )^{2}}  \label{A.4} \\
b^{(0)}(q^{2};m_{1}^{2},m_{2}^{2}) &=&\frac{1}{(4\pi )^{2}}\bigg\{%
2-\Lg(m_{1}^{2})+\sum_{j=1}^{2}x_{j}\log (1-\frac{1}{x_{j}})\bigg\}  \notag \\
b^{(1)}(q^{2};m_{1}^{2},m_{2}^{2}) &=&\frac{1}{(4\pi )^{2}}\{4+\frac{\pi ^{2}%
}{12}-2\Lg(m_{1}^{2})+\frac{\Lg(m_{1}^{2})^{2}}{2}  \notag \\
&&+\frac{1}{2}\sum_{j=1}^{2}x_{j}\log (1-\frac{1}{x_{j}})[4-2\Lg(q^{2}-i0)-%
\log (1-x_{j})-\log (-x_{j})]  \notag \\
&&-x_{1}[\log (1-x_{1})\log (1-x_{2})-\log (-x_{1})\log (-x_{2})]  \notag \\
&&+(x_{1}-x_{2})[\log (x_{2}-x_{1})\log (1-\frac{1}{x_{2}})
- \text{Li}_{2}(\frac{%
1-x_{2}}{x_{1}-x_{2}})+ \text{Li}_{2}(\frac{-x_{2}}{x_{1}-x_{2}})]\} \notag
\end{eqnarray}
where $D=4-2\varepsilon$ is the dimension of space-time,
$x_{1/2}$ are given by
\begin{equation}
x_{1,2} \;=\; \frac{1}{2q^2} \bigg\{ m_2^2 - m_1^2 + q^2 \pm
\sqrt{\lambda(m_1^2,m_2^2,q^2)} \,\bigg\} \:,
\end{equation}
$\lambda$ being the K{\"a}ll{\'e}n-function
$\lambda(a,b,c) = (a-b-c)^2-4bc$,
and where we have expanded to order $\varepsilon^1$. All masses carry an infinitesimal negative
imaginary part.

Tensor integrals can be reduced to scalar ones by decomposing them with
respect to Lorentz covariants. The notation is
\begin{align}
A_{r}(m^{2}) \;=\; & \; \text{coefficient of the tensor
integral with }r\text{ momenta }k \\
&\;\text{in the numerator ($r$ even)}  \notag \\
B_{rs}(q^{2};m_{1}^{2},m_{2}^{2}) \;=\; &
\;\text{coefficient of the tensor
integral with }r\text{ momenta }k  \label{A.5} \\
&\;\text{in the numerator, and }s\text{ factors of }g_{\mu \nu }\text{ on the
rhs,}  \notag
\end{align}%
\ e.g.
\begin{equation}
\mu ^{4-D}\int \frac{d^{D}k}{i(2\pi )^{D}}\;\frac{k^\mu k^\nu}{%
k^{2}-m^{2}} \;=\; g^{\mu\nu} \: A_2  \label{A2_definition}
\end{equation}
and
\begin{eqnarray}
\mu ^{4-D}\int \frac{d^{D}k}{i(2\pi )^{D}}\;\frac{k^{\mu }}{%
[(k+q)^{2}-m_{1}^{2}][k^{2}-m_{2}^{2}]} &=&q^{\mu }B_{10}  \label{A.6} \\
\mu ^{4-D}\int \frac{d^{D}k}{i(2\pi )^{D}}\;\frac{k^{\mu }\,k^{\nu }}{%
[(k+q)^{2}-m_{1}^{2}][k^{2}-m_{2}^{2}]} &=&q^{\mu }q^{\nu }B_{20}+g^{\mu \nu
}B_{21}  \notag \\
\mu ^{4-D}\int \frac{d^{D}k}{i(2\pi )^{D}}\;\frac{k^{\mu }\,k^{\nu
}\,k^{\rho }}{[(k+q)^{2}-m_{1}^{2}][k^{2}-m_{2}^{2}]} &=&q^{\mu }q^{\nu
}q^{\rho }B_{30}+(g^{\mu \nu }q^{\rho }+g^{\mu \rho }q^{\nu }+g^{\nu \rho
}q^{\mu })B_{31}.  \notag
\end{eqnarray}
$A_2$ is given in terms of $A_0$ in eq.~(\ref{A0_and_A2}), and
the functions $B_{rs}$ can all be related to $B_{0}$ (and $A_{0}$). For the
form factor $f_+$ we need
\begin{eqnarray}
B_{21}(q^{2};m_{1}^{2},m_{2}^{2}) &=& \frac{1}{4q^{2}(1-D)} \{
(m_{2}^{2}-m_{1}^{2}-q^{2})A_{0}(m_{1}^{2})
\label{A.7} \\
&& \hphantom{\frac{1}{4q^{2}(1-D)} \{}
+ \; (m_{1}^{2}-m_{2}^{2}-q^{2})A_{0}(m_{2}^{2})
\notag\\
&& \hphantom{\frac{1}{4q^{2}(1-D)} \{}
+\;\lambda
(q^{2},m_{1}^{2},m_{2}^{2})B_{0}(q^{2};m_{1}^{2},m_{2}^{2})\}  \notag \\
B_{31}(q^{2};m_{1}^{2},m_{2}^{2}) &=&\frac{1}{8q^{4}(1-D)} \{
[q^{4}-m_{1}^{4}-m_{2}^{4}+2m_{1}^{2}m_{2}^{2}+4m_{1}^{2}q^{2}-\frac{%
4m_{1}^{2}q^{2}}{D}]A_{0}(m_{1}^{2})  \notag \\
&&\hspace{5.5em}+\;[\lambda (q^{2},m_{1}^{2},m_{2}^{2})+\frac{4m_{2}^{2}q^{2}%
}{D}]A_{0}(m_{2}^{2})  \notag \\
&&\hspace{5.5em}+\;(m_{1}^{2}-m_{2}^{2}-q^{2})\lambda
(q^{2},m_{1}^{2},m_{2}^{2})B_{0}(q^{2};m_{1}^{2},m_{2}^{2}) \}.
\end{eqnarray}

In addition, there are 1-loop integrals which involve higher powers of propagators.
These are obtained from the formulae above by differentiation with respect to $m^{2}$:
\begin{eqnarray}
A_{0;n}(m^{2}) &=&\frac{1}{(n-1)!}\,[\frac{d}{dm^{2}}]^{n-1}A_{0}(m^{2})
\label{A02_deriv_definition} \\
B_{rs;\;n_{1} n_{2}}(q^{2};m_{1}^{2},m_{2}^{2}) &=&\frac{1}{%
(n_{1}-1)!\,(n_{2}-1)!}\,[\frac{\partial}{\partial m_{1}^{2}}]^{n_{1}-1}
[\frac{\partial}{\partial m_{2}^{2}}]^{n_{2}-1}B_{rs}(q^{2};m_{1}^{2},m_{2}^{2})\;.\qquad  \label{6.8}
\end{eqnarray}


\section{Reducible contributions to the $K_{\ell 3}$ form factor $f_{+}$}
\label{redContribToFPlus}

In this appendix we give the reducible contribution $\Delta $ of each
diagram for the $K_{\ell 3}$ form factor $f_{+}$. The upper index of $%
\Delta $ refers to a specific diagram in fig.~2.

The basic 1-loop functions occuring here (and in the other form factor
$f_-$) are $A_0$ and $B_0$, and tensors and mass derivatives thereof,
cf. appendix \ref{OneLoopIntegrals}.

{\scriptsize
\begin{eqnarray}
\Diag{0}{}{K\pi}{+} &=& 1
\\
\Diag{1a}{}{K\pi}{+} &=&
-\frac{3}{2\vierpi{2} F^2} \; \Big\{
\BOne{2}{1}{q^2}{m_\eta^2}{m_K^2}
+ \BOne{2}{1}{q^2}{m_K^2}{m_\pi^2}
\Big\}
\\
\Diag{1b}{}{K\pi}{+} &=&
\frac{1}{6 \vierpi{2} F^2} \; \Big\{
3 A_0(m_\eta^2) + 7 A_0(m_K^2) + 5 A_0(m_\pi^2)
\Big\}
\\
\Diag{1c}{}{K\pi}{+} &=&
\frac{2}{F^2} \; \Big\{
4 L_4 (m_\pi^2 + 2 m_K^2) + 2 L_5 (m_\pi^2 + m_K^2) + q^2 L_9
\Big\}
\\
\Diag{2a}{}{K\pi}{+} &=&
\frac{2}{\vierpi{2} F^4} \; \Big\{
2 \BOne{3}{1}{q^2}{m_\eta^2}{m_K^2} \; L_3 (m_\pi^2 - m_K^2)
\\
&&\quad
- 2 \BOne{3}{1}{q^2}{m_K^2}{m_\pi^2} \;
[8 L_1 + 4 L_2 + L_3] (m_\pi^2 - m_K^2)
\nonumber\\
&&\quad
+ \BOne{2}{1}{q^2}{m_\eta^2}{m_K^2}  \:\big[ L_3 (m_\pi^2 - m_K^2 + 3 q^2)
- 6 L_4 (m_\pi^2 + 2 m_K^2) - 2 L_5 (m_\pi^2 + 5 m_K^2) \big]
\nonumber\\
&&\quad
- \BOne{2}{1}{q^2}{m_K^2}{m_\pi^2} \:\big[
8 L_1 (m_\pi^2 - m_K^2 - q^2)
+ 4 L_2 (m_\pi^2 - m_K^2 + q^2)
\nonumber\\
&&\hspace{7em}
+ L_3 (m_\pi^2 - m_K^2 - 3 q^2)
+ 2 L_4 (7 m_\pi^2 + 10 m_K^2)
+ 6 L_5 (m_\pi^2 + m_K^2) \big]
\Big\}\notag
\\
\Diag{2b}{}{K\pi}{+} &=&
\frac{1}{\vierpi{2} F^4} \; \Big\{
3 \BOne{2}{1}{q^2}{m_K^2}{m_\pi^2} \:\big[
- 4 L_4 (m_\pi^2 + 2 m_K^2) - 2 L_5 (m_\pi^2 + m_K^2) - q^2 L_9 \big]
\\
&&\quad
+ \BOne{2}{1}{q^2}{m_\eta^2}{m_K^2}  \:\big[
- 12 L_4 (m_\pi^2 + 2 m_K^2) + 2 L_5 (m_\pi^2 - 7 m_K^2) - 3\, q^2 L_9 \big]
\Big\}\notag
\\
\Diag{2c}{}{K\pi}{+} &=&
\frac{1}{3 F^4} \Big\{
\\&&\hspace{-5.5em}
A_0(m_\eta^2) \: [
-48 L_1 m_\eta^2
-14 L_3 m_\eta^2
+4 L_4 (14 m_K^2+m_\pi^2)
+13 L_5 (2 m_K^2+m_\pi^2)
+3 q^2 L_9]
\notag\\&&\hspace{-6.5em}
+A_0(m_K^2) \: [
-192 L_1 m_K^2
-24 L_2 m_K^2
-60 L_3 m_K^2
+4 L_4 (44 m_K^2+7 m_\pi^2)
+2 L_5 (31 m_K^2+7 m_\pi^2)
+7 q^2 L_9 ]
\nonumber\\&&\hspace{-6.5em}
+A_0(m_\pi^2) \: [
-144 L_1 m_\pi^2
-24 L_2 m_\pi^2
-54 L_3 m_\pi^2
+4 L_4 (10 m_K^2+29 m_\pi^2)
+2 L_5 (2 m_K^2+23 m_\pi^2)
+5 q^2 L_9]
\nonumber\\&&\hspace{-6.5em}
-2 \: A_2(m_\eta^2) \: [24 L_2+5 L_3]
-12 \: A_2(m_K^2) \: [4 L_1+18 L_2+5 L_3]
-6 \: A_2(m_\pi^2) \: [8 L_1+28 L_2+7 L_3]
\Big\}\nonumber
\\
\Diag{2d}{}{K\pi}{+} &=&
\frac{1}{F^4}  \; \Big\{
12 \: \BOne{2}{1}{q^2}{m_\eta^2}{m_K^2}
\: [L_4 (2 m_K^2 + m_\pi^2) + L_5 m_K^2]
\\
&&\qquad
+4 \: \BOne{2}{1}{q^2}{m_K^2}{m_\eta^2}
\: [3 L_4 (2 m_K^2 + m_\pi^2) + L_5 (4 m_K^2 - m_\pi^2)]
\notag\\
&&\qquad
+12 \: \BOne{2}{1}{q^2}{m_K^2}{m_\pi^2}
\: [L_4 (2 m_K^2 + m_\pi^2) + L_5 m_\pi^2]
\notag\\
&&\qquad
+12 \: \BOne{2}{1}{q^2}{m_\pi^2}{m_K^2}
\: [L_4 (2 m_K^2 + m_\pi^2) + L_5 m_K^2]
\notag\\
&&\hspace{-5em}
+12 \: \BOneNenner{2}{1}{q^2}{1}{m_\eta^2}{2}{m_K^2}
    \: m_K^2 \:[L_4 (2 m_K^2 + m_\pi^2) + L_5 m_K^2
    -2 L_6 (2 m_K^2 + m_\pi^2) - 2 L_8 m_K^2 ]
\notag\\
&&\hspace{-5em}
+12 \: \BOneNenner{2}{1}{q^2}{1}{m_K^2}{2}{m_\pi^2}
   \: m_\pi^2 \:[
   L_4 (2 m_K^2 + m_\pi^2) + L_5 m_\pi^2
   -2 L_6 (2 m_K^2 + m_\pi^2) - 2 L_8 m_\pi^2 ]
\notag\\
&&\hspace{-5em}
+12 \: \BOneNenner{2}{1}{q^2}{1}{m_\pi^2}{2}{m_K^2}
   \: m_K^2 \:[
   L_4 (2 m_K^2 + m_\pi^2) + L_5 m_K^2
   -2 L_6 (2 m_K^2 + m_\pi^2) - 2 L_8 m_K^2]
\notag\\
&&\hspace{-5em}
+4 \: \BOneNenner{2}{1}{q^2}{1}{m_K^2}{2}{m_\eta^2}
   \: [3 L_4 m_\eta^2 (2 m_K^2 + m_\pi^2)
   + L_5 m_\eta^2 ( 4 m_K^2 - m_\pi^2)
   -2 L_6 (8 m_K^4 + 2 m_K^2 m_\pi^2 - m_\pi^4)
\notag\\
&&\hspace{5em}
   -16 L_7 (m_K^2 - m_\pi^2)^2
   -2 L_8 (8 m_K^4 - 8 m_K^2 m_\pi^2 + 3 m_\pi^4)]
\Big\}\notag
\\
\Diag{2e}{}{K\pi}{+} &=&
\frac{1}{3 F^4}  \; \Big\{
-4 A_0(m_\eta^2) \: [3 L_4 (2 m_K^2 + m_\pi^2) + L_5 (4 m_K^2 - m_\pi^2)]
\\
&&\qquad\quad
-28 A_0(m_K^2) \: [L_4 (2 m_K^2 + m_\pi^2) + L_5 m_K^2 ]
\notag\\
&&\qquad\quad
-20 A_0(m_\pi^2) \: [L_4 (2 m_K^2 + m_\pi^2) + L_5 m_\pi^2]
\notag\\
&&\hspace{-1em}
-28 \ANenner{2}{m_K^2} \: m_K^2 \: [
    L_4 (2 m_K^2 + m_\pi^2) + L_5 m_K^2
    -2 L_6 (2 m_K^2 + m_\pi^2) - 2 L_8 m_K^2 ]
\notag\\
&&\hspace{-1em}
-20 \ANenner{2}{m_\pi^2} \: m_\pi^2 \: [
    L_4 (2 m_K^2 + m_\pi^2) + L_5 m_\pi^2
    -2 L_6 (2 m_K^2 + m_\pi^2) - 2 L_8 m_\pi^2 ]
\notag\\
&&\hspace{-1em}
-4 \ANenner{2}{m_\eta^2} \: [
   3 L_4 m_\eta^2 (2 m_K^2 + m_\pi^2)
   + L_5 m_\eta^2 (4 m_K^2 - m_\pi^2)
   -2 L_6 (8 m_K^4 + 2 m_K^2 m_\pi^2 - m_\pi^4)
\notag\\
&&\hspace{5em}
   -16 L_7 (m_K^2 - m_\pi^2)^2
   -2 L_8 (8 m_K^4 - 8 m_K^2 m_\pi^2 + 3 m_\pi^4)]
\Big\}\notag
\end{eqnarray}
\begin{eqnarray}
\Diag{3a}{}{K\pi}{+} &=&
\frac{9}{4\vierpi{4} F^4} \; \Big\{
 \BOne{2}{1}{q^2}{m_\eta^2}{m_K^2} ^2
+ 2 \BOne{2}{1}{q^2}{m_\eta^2}{m_K^2} \: \BOne{2}{1}{q^2}{m_K^2}{m_\pi^2}
\\
&&\quad
+ \BOne{2}{1}{q^2}{m_K^2}{m_\pi^2}^2
\Big\}\notag
\\
\Diag{3b}{}{K\pi}{+} &=&
- \frac{1}{4\vierpi{4} F^4} \; \Big\{
 3 A_0(m_\eta^2) \: \BOne{2}{1}{q^2}{m_\eta^2}{m_K^2}
+ 3 A_0(m_\eta^2) \: \BOne{2}{1}{q^2}{m_K^2}{m_\pi^2}
\\
&&\qquad
+ 9 A_0(m_K^2) \: \BOne{2}{1}{q^2}{m_\eta^2}{m_K^2}
+ 7 A_0(m_K^2) \: \BOne{2}{1}{q^2}{m_K^2}{m_\pi^2}
\nonumber\\
&&\qquad
+ 3 A_0(m_\pi^2) \: \BOne{2}{1}{q^2}{m_\eta^2}{m_K^2}
+ 5 A_0(m_\pi^2) \: \BOne{2}{1}{q^2}{m_K^2}{m_\pi^2}
\Big\}\notag
\\
\Diag{3c}{}{K\pi}{+} &=&
\frac{1}{24 F^4} \: \Big\{
9 \: \BOne{2}{1}{q^2}{m_\eta^2}{m_K^2} \: [A_0(m_\pi^2) + 6 \, A_0(m_K^2) + A_0(m_\eta^2) ]
\\
&&\qquad\quad
+ \BOne{2}{1}{q^2}{m_\pi^2}{m_K^2} \: [33 \, A_0(m_\pi^2) + 30 \, A_0(m_K^2) + 9 \, A_0(m_\eta^2) ]
\notag\\
&&\qquad\quad
-6 \: \BOneNenner{2}{1}{q^2}{1}{m_K^2}{2}{m_\pi^2} \: m_\pi^2 \: [ A_0(m_\eta^2) - 3 \, A_0(m_\pi^2) ]
\notag\\
&&\qquad\quad
+3 \: \BOneNenner{2}{1}{q^2}{1}{m_\pi^2}{2}{m_K^2} \: A_0(m_\eta^2) (3 m_\eta^2+m_\pi^2)
\notag\\
&&\qquad\quad
+3 \: \BOneNenner{2}{1}{q^2}{1}{m_\eta^2}{2}{m_K^2} \: A_0(m_\eta^2) (3 m_\eta^2+m_\pi^2)
\notag\\
&&\hspace{-1em}
- 2\, \BOneNenner{2}{1}{q^2}{1}{m_K^2}{2}{m_\eta^2} \:
  [ 9 m_\pi^2 \, A_0(m_\pi^2) - 6 (3 m_\eta^2+m_\pi^2) \: A_0(m_K^2)
    + (16 m_K^2-7 m_\pi^2) \: A_0(m_\eta^2)
\Big\}\notag
\\
\Diag{3d}{}{K\pi}{+} &=&
-\frac{1}{72 F^4}  \; \Big\{
\\&&\qquad
57 \: A_0(m_\eta^2) \: A_0(m_K^2)
+42 \: A_0(m_K^2)^2
+41 \: A_0(m_K^2) \: A_0(m_\pi^2)
+40 \: A_0(m_\pi^2)^2
\notag\\
&&\qquad
-2 \: \ANenner{2}{m_\eta^2} \: A_0(m_\eta^2) \: (16 m_K^2-7 m_\pi^2)
+12 \: \ANenner{2}{m_\eta^2} \: A_0(m_K^2) \: (3 m_\eta^2+m_\pi^2)
\notag\\
&&\qquad
-18 \: \ANenner{2}{m_\eta^2} \: A_0(m_\pi^2) \: m_\pi^2
+7 \: \ANenner{2}{m_K^2} \: A_0(m_\eta^2) \: (3 m_\eta^2+m_\pi^2)
\notag\\
&&\qquad
-10 \: \ANenner{2}{m_\pi^2} \: A_0(m_\eta^2) \: m_\pi^2
+30 \: \ANenner{2}{m_\pi^2} \: A_0(m_\pi^2) \: m_\pi^2
\Big\}\notag
\\
\Diag{3e}{}{K\pi}{+} &=&
- \frac{1}{18\vierpi{4} F^4} \; \Big\{
 9 A_0(m_\eta^2) \: \BOne{2}{1}{q^2}{m_\eta^2}{m_K^2}
+ 6 A_0(m_\eta^2) \: \BOne{2}{1}{q^2}{m_K^2}{m_\pi^2}
\\
&&\qquad\qquad
+ 24 A_0(m_K^2) \: \BOne{2}{1}{q^2}{m_\eta^2}{m_K^2}
+ 19 A_0(m_K^2) \: \BOne{2}{1}{q^2}{m_K^2}{m_\pi^2}
\nonumber\\
&&\qquad\qquad
+ 12 A_0(m_\pi^2) \: \BOne{2}{1}{q^2}{m_\eta^2}{m_K^2}
+ 20 A_0(m_\pi^2) \: \BOne{2}{1}{q^2}{m_K^2}{m_\pi^2}
\Big\}\notag
\\
\Diag{3f}{}{K\pi}{+} &=& 
\frac{1}{360 F^4} \Big\{
81 \: A_0(m_\eta^2)^2
+168 \: A_0(m_\eta^2) \: A_0(m_K^2)
+132 \: A_0(m_\eta^2) \: A_0(m_\pi^2)
\\&&\qquad\qquad
+318 \: A_0(m_K^2)^2
+386 \: A_0(m_K^2) \: A_0(m_\pi^2)
+175 \: A_0(m_\pi^2)^2
\Big\}\nonumber
\end{eqnarray}
} 


\section{Divergent parts of $\mathcal{L}^{(6)}$-constants}
\label{divPartsOfBeta}

In this appendix we list the divergent parts of all $\mathcal{L}^{(6)}$-constants
which occur in the meson vector form factors.
They are derived from eqs.~(\ref{5.19a}), (\ref{5.19b}) on the one hand and
eqs.~(\ref{5.22}), (\ref{5.23})
on the other hand taking into consideration eq.~(\ref{5.21}).
Similar relations for the electromagnetic form factors of $\pi^\pm$,
$K^\pm$, $K_0$, and the weak form factors $f_\pm$ for the $\eta\to K^\pm W^\mp$ 
decay are also taken into consideration
\cite{PostSchilcher}\cite{dissPeter}.

Let
\begin{equation}
  \beta_j \;=\; \frac{\beta_j^{(-2)}}{\varepsilon^2} +
  \frac{\beta_j^{(-1)}}{\varepsilon} + \beta_j^{(0)}
  + \mathcal{O}(\varepsilon)
\end{equation}
the Laurent expansion of the $\mathcal{L}^{(6)}$-parameters,
cf. eq.~(\ref{betaGL}). For their divergent parts
\begin{equation}
  \beta_j^{\text{\it div}} \;=\; \frac{\beta_j^{(-2)}}{\varepsilon^2} +
  \frac{\beta_j^{(-1)}}{\varepsilon}
\end{equation}
we find
\begin{eqnarray}
\beta_{22}^{\text{\it div}}+\beta_{23}^{\text{\it div}}
&=& \frac{1}{(4\pi)^2} \bigg\{
    \frac{1}{64(4\pi)^2 \!\;\varepsilon^2}
    \\&&\qquad
    - \; \frac{1}{\!\;\varepsilon} \Big[ \frac{1}{96(4\pi)^2}
    -\frac{2}{3} \Lren{1} + \frac{1}{3} \Lren{2} - \frac{1}{2} \Lren{3} + \frac{1}{4} \Lren{9}
    \Big] \bigg\} \notag
\\
2 \beta_{24}^{\text{\it div}}-\beta_{25}^{\text{\it div}}
&=& -\frac{1}{(4\pi)^2} \frac{3}{2\!\;\varepsilon} \Lren{3}
\\
\beta_{26}^{\text{\it div}}
&=& \frac{1}{(4\pi)^2} \bigg\{
    \frac{1}{32(4\pi)^2\!\;\varepsilon^2}
    +\frac{1}{\varepsilon} \Big[
    \frac{7}{192(4\pi)^2}-\frac{1}{2} \Lren{3}-\frac{1}{2} \Lren{9}
    \Big] \bigg\}
\\
\beta_{22}^{\text{\it div}} - 2 \beta_{24}^{\text{\it div}}
+ \beta_{27}^{\text{\it div}}
&=& \frac{1}{(4\pi)^2} \bigg\{
       \frac{-11}{96(4\pi)^2\!\;\varepsilon^2}
       - \frac{1}{\varepsilon} \Big[
       \frac{83}{1152(4\pi)^2}-2 \Lren{1}+\Lren{2}
       \\&&\qquad
       - \; \frac{3}{2} \Lren{3}
       - \frac{2}{3} \Lren{4}-\frac{1}{2} \Lren{5}
       -\frac{3}{4} \Lren{9} \Big] \bigg\} \notag
\\
\beta_{14}^{\text{\it div}}
&=& \frac{1}{(4\pi)^2} \bigg\{
       \frac{-7}{128(4\pi)^2\!\;\varepsilon^2}
       +\frac{1}{\varepsilon} \Big[
       \frac{35}{1152(4\pi)^2} + \frac{1}{3} \Lren{1} + \frac{1}{6} \Lren{2}
       \\&&\qquad
       + \; \frac{1}{36} \Lren{3} - \Lren{4} + \frac{3}{4} \Lren{5}
       \Big] \bigg\} \notag
\\
\beta_{8}^{\text{\it div}}
&=& \frac{1}{(4\pi)^2} \bigg\{
      \frac{-19}{64(4\pi)^2\!\;\varepsilon^2}
      -\frac{1}{\varepsilon} \Big[
      \frac{13}{384(4\pi)^2}-\frac{4}{3} \Lren{1} - \frac{8}{3} \Lren{2}
      \\&&\qquad
      - \; \frac{16}{9} \Lren{3} - \frac{3}{2} \Lren{5}
      \Big] \bigg\} \notag
\\
\beta_{16}^{\text{\it div}}+\beta_{18}^{\text{\it div}}
&=& \frac{1}{(4\pi)^2} \bigg\{
       \frac{3}{64(4\pi)^2\!\;\varepsilon^2}
       \\&&\qquad
       - \; \frac{1}{\varepsilon} \Big[
       \frac{119}{384(4\pi)^2} + \frac{4}{3} \Lren{1}
       +\frac{14}{3} \Lren{2}
       + \frac{28}{9} \Lren{3}
       +5 \Lren{4}
       \notag\\&&\qquad
       - \; \frac{37}{18} \Lren{5}
       -6 \Lren{6}
       + 8 \Lren{7} - 2 \Lren{8}
       \Big] \bigg\} \notag
\\
\beta_{17}^{\text{\it div}}
&=& \frac{1}{(4\pi)^2} \bigg\{
       \frac{-17}{128(4\pi)^2\!\;\varepsilon^2}
       +\frac{1}{\varepsilon} \Big[
       \frac{173}{768(4\pi)^2}
       + \frac{2}{3} \Lren{1}
       \\&&\qquad
       + \; \frac{7}{3} \Lren{2}
       +\frac{11}{9} \Lren{3}
       +\Lren{4}
       - \frac{13}{12} \Lren{5}
       +12 \Lren{7}
       +3 \Lren{8}
       \Big] \bigg\} \notag
\\
\beta_{15}^{\text{\it div}}
&=& \frac{1}{(4\pi)^2} \bigg\{
       \frac{7}{192(4\pi)^2\!\;\varepsilon^2}
       -\frac{1}{\varepsilon} \Big[
       \frac{5}{324(4\pi)^2} + \frac{2}{9} \Lren{1}
       \\&&\qquad
       + \; \frac{1}{9} \Lren{2} + \frac{1}{54} \Lren{3}
       - \frac{2}{3} \Lren{4}
       + \frac{1}{2} \Lren{5}
       \Big] \bigg\} \notag
\\
\beta_{20}
&=& \frac{1}{(4\pi)^2} \bigg\{
    \frac{-11}{384(4\pi)^2\!\;\varepsilon^2}
    -\frac{1}{\varepsilon} \Big[
    \frac{125}{2304(4\pi)^2}
    +\frac{2}{9} \Lren{1}
    \\&&\quad
    + \; \frac{7}{9} \Lren{2}
    +\frac{17}{108} \Lren{3}
    +\frac{1}{3} \Lren{4}
    -\frac{17}{12} \Lren{5}
    +4 \Lren{7}+2 \Lren{8}
    \Big] \bigg\} \:. \notag
\end{eqnarray}
where $L_i^{(0)}$ are the $\varepsilon^0$-coefficients of the
$\mathcal{L}^{(4)}$-parameters, cf. eq.~(\ref{5.41b}).

Note that the $\beta_j$ do not contain any mass terms. Thus, the only source
of masses in the Lagrangian (\ref{3.9}) is the mass matrix $\chi$
defined in (\ref{3.3}).


\section{Electromagnetic pion form factor}
\label{pionFFdiscussion}

The $\mathcal{O}(p^{6})$ calculation of the pion electromagnetic form factor is
carried out in a manner completely analogous to that of the kaon
semi-leptonic form factors. We quote the result of the $\mathcal{L}^{(6)}$
contribution:
\begin{eqnarray}
F^{\pi ^{+}}[\mathcal{L}^{(6)}] &=& \hphantom{+\;}
 \frac{2q^{4}}{F^{4}}\;\{\beta_{22}+\beta_{23}\} \label{8.1} \\
&&+\;\frac{2q^{2}}{F^{4}}\;\{-2m_{\pi}^{2}\beta_{22}+2m_{\pi}^{2}\beta_{24}+m_{\pi
}^{2}\beta_{25}+(m_{\pi}^{2}+2m_{K}^{2})\beta_{26}-2m_{\pi}^{2}\beta_{27}\} \:. \notag
\end{eqnarray}
For the interpolation polynomials of the $\mathcal{O}(p^{6})$ loop corrections
(reducible plus irreducible diagrams) we obtain
\begin{equation}
\Delta_{p^{6}}^{loop}[F^{\pi ^{+}}]
\;=\; x(0.0813+0.0619\,x+0.0387\,x^{2}+0.0218\,x^{3}+0.0068\,x^{4})\qquad
\label{8.2}
\end{equation}
for $x = q^2/m_K^2 \in [-1,(2m_\pi)^2/m_K^2]$) and in the decay region
\begin{align}
\Delta_{p^{6}}^{loop}[F^{\pi^{+}}]
\;=\;&
   - 0.0084
   + 0.1076 \, x
   + 0.0956 \, x^2
   - 0.0544 \, x^3
   + 0.0233 \, x^4
\\&
   -\; 0.0047 \, x^{5}
   + i \, (
     0.0120
   - 0.0998 \, x
   + 0.2435 \, x^2
   + 0.1848 \, x^3
\notag\\
&
+\;  0.1087 \, x^4
   + 0.0260 \, x^{5} )
\notag
\end{align}
for $x = q^2/m_K^2 \in [4m_\pi^2/m_K^2,1]$).

The separate reducible and irreducible $\mathcal{O}(p^{6})$ loop contributions (modulo
a quadratic polynomial) to the pion form factor are plotted in fig.~6.

The arbitrary linear term can be fitted by using the experimental pion charge
radius \cite{AmendoliaPIC}%
\begin{equation}
\left\langle r^{2}\right\rangle^{\pi^{+}} \;=\;
(0.439\pm 0.008)\;\text{fm}^{2}\;.
\label{8.3}
\end{equation}
The constant multiplying $q^{4}$ is obtained by using the curvature
of the form factor at the origin. We chose a value of $3.4\:\text{GeV}^{-4}$,
which represents an average of some typical extractions from experiment
\cite{GasserMeissner}, \cite{ErkalOlsson}, \cite{ColFinkUrech}.
As fig.~7 demonstrates this choice of parameters yields
a good description of the experimental form factor.


\begin{figure}
  \centerline{
    \includegraphics*[width=5.7553in]{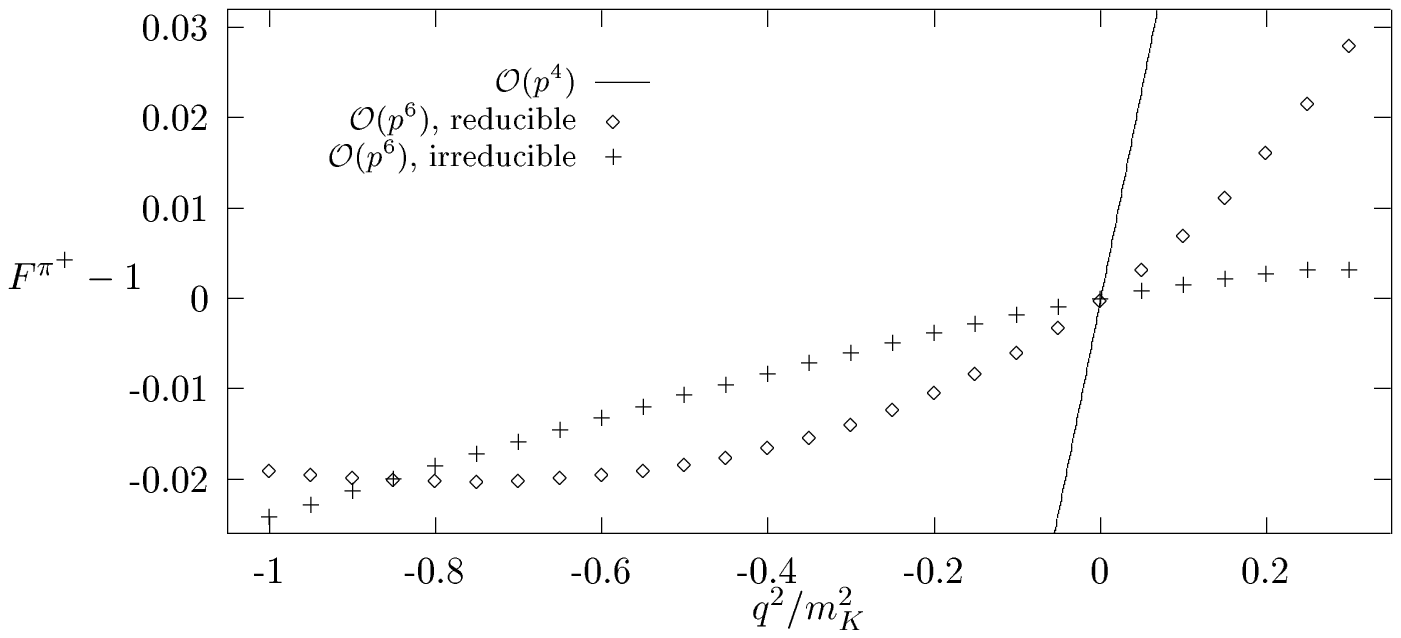}
  }
  Figure 6:
  \footnotesize
  $\mathcal{O}(p^{6})$ loop contributions to the pion electromagnetic
  form factor, in analogy to fig.~3.
\end{figure}


\begin{figure}
  \centerline{
    \includegraphics*[width=5.6853in]{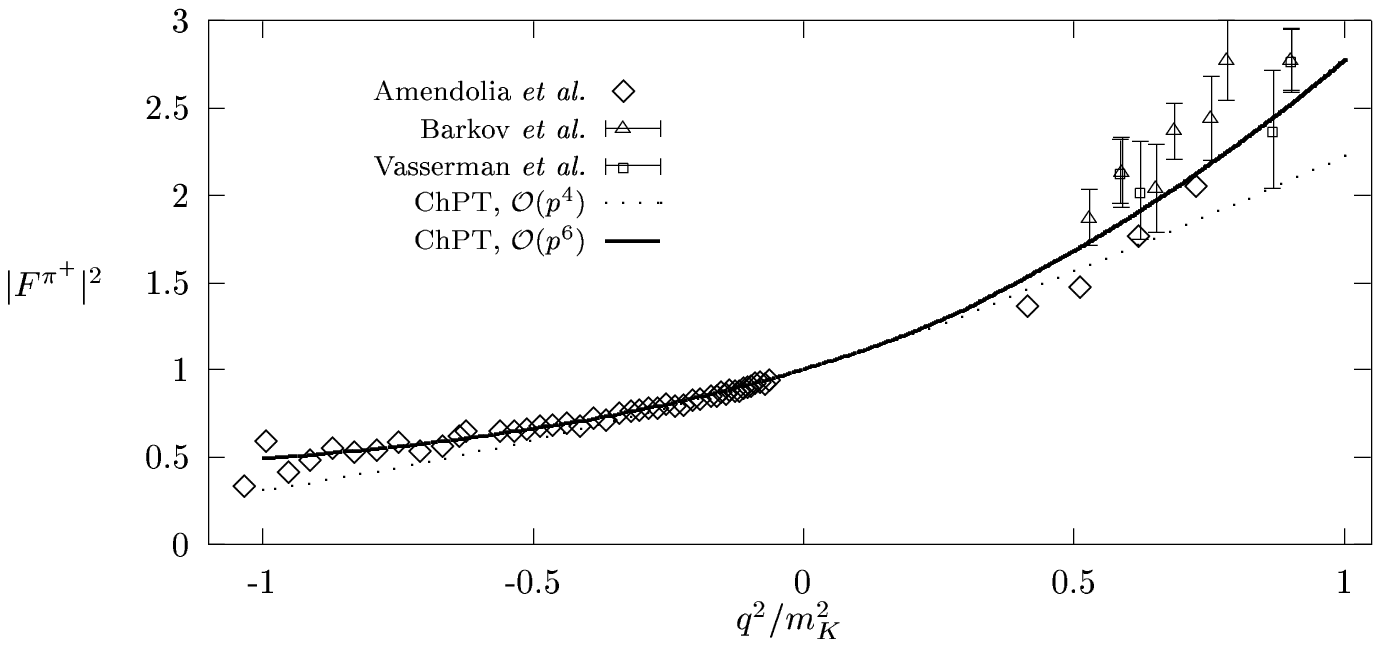}
  }
  Figure 7:
  \footnotesize
  The pion electromagnetic form factor in $\mathcal{O}(p^{4})$
  and $\mathcal{O}(p^{6})$ chiral perturbation theory versus experiment.
  The data are from ref. \protect\cite{AmendoliaPIC},
  \protect\cite{BarkovPIC}, and \protect\cite{VassermannPIC}.
\end{figure}

\end{appendix}

\vspace{6mm}

\noindent
{\Large \textbf{Acknowledgements}} \\

\noindent
We thank G. Colangelo for critical comments.

\clearpage



\end{document}